%% file: main.tex
\documentclass[prb,twocolumn,amsmath,amssymb,superscriptaddress]{revtex4-2}
\usepackage{epsfig, graphicx,graphics,amsmath,amssymb,float}
\usepackage[T1]{fontenc}
\usepackage[latin9]{inputenc}
\usepackage{amsmath}
\usepackage{amssymb}
\usepackage{appendix}
\usepackage{amscd}
\usepackage{bm}
\usepackage{psfrag}
\usepackage{bbm} 
\usepackage{babel}
\usepackage{wasysym }
\usepackage{mathrsfs}
\usepackage{color}
\usepackage[normalem]{ulem}
\usepackage[final]{hyperref} 
\hypersetup{
	colorlinks=true,       
	linkcolor=blue,        
	citecolor=blue,        
	filecolor=magenta,     
	urlcolor=blue         
}

\usepackage[american,siunitx]{circuitikz}
\usepackage{amsmath}
\usetikzlibrary{shapes,decorations.pathmorphing}

\begin{document}

\title{Error threshold in active steering protocols for few-qubit systems}

\author{Nico Ackermann}
\affiliation{
Departamento de F{\'i}sica Te{\'o}rica de la Materia Condensada, Condensed Matter Physics Center (IFIMAC)
and Instituto Nicol{\'a}s Cabrera, 
Universidad Aut{\'o}noma de Madrid, 28049 Madrid, Spain}

\author{Samuel Morales}
\affiliation{Institut f\"ur Theoretische Physik, Heinrich-Heine-Universit\"at, D-40225  D\"usseldorf, Germany}

\author{Alfredo Levy Yeyati}
\affiliation{
Departamento de F{\'i}sica Te{\'o}rica de la Materia Condensada, Condensed Matter Physics Center (IFIMAC)
and Instituto Nicol{\'a}s Cabrera, 
Universidad Aut{\'o}noma de Madrid, 28049 Madrid, Spain}

\author{Sebastian Diehl}
\affiliation{Institut f\"ur Theoretische Physik, Universit\"at zu K\"oln, Z\"ulpicher Stra{\ss}e 77, 50937 Cologne, Germany}

\author{Reinhold Egger}
\affiliation{Institut f\"ur Theoretische Physik, Heinrich-Heine-Universit\"at, D-40225  D\"usseldorf, Germany}

\begin{abstract}
We study active steering protocols for weakly measured qubits in the presence of error channels due to 
amplitude and phase noise. If the error rate is sufficiently small, the protocol approaches and 
stabilizes a predesignated pure target state with high fidelity and high purity, and thus implements 
autonomous state stabilization. 
We present numerical simulation results for one and two qubits,  taking Andreev qubit circuits as example. 
As function of the error rate, a sharp threshold separates an error-correcting weak-damping regime from a
strong-damping regime where the target state cannot be reached anymore. At the threshold, the purity gap closes.
\end{abstract}
\maketitle

\section{Introduction}\label{sec1}

Engineering and protecting quantum states from the detrimental effects of errors due to decoherence or noise is a fundamental ingredient for any quantum information processing scheme. A promising avenue towards robust state preparation uses dissipation as a resource to target quantum states of interest \cite{Poyatos1996,Diehl2008,Verstraete2009,Barreiro_2011,Krauter2011}. On the other hand, quantum error correction (QEC) employs the feedback from repeated projective syndrome measurements in stabilizer codes \cite{Nielsen2000,Terhal2015,Brown2016}, which has seen recent experimental breakthroughs \cite{Sivak2023,Ni2023,Krinner2022,Acharya2023}.  
Recent work combines these threads, and aims at the ``passive'' protection of quantum states in suitably engineered driven-dissipative systems \cite{Paz1998,Barnes2000,Ahn2002,Ahn2003,Sarovar2004,Oreshkov2007,Wiseman2010,Kerckhoff2010,Kapit2016,Kapit2018,Gau2020a,Gau2020b,Lieu2020,Lieu2023,Shtanko2023,Kristensen2023}. 
Related ideas have been implemented experimentally in few-qubit systems, see, e.g., Refs.~\cite{Leghtas2013,Minev2019,Campagne2020,Gertler2021,Livingston2022,Lachance2024}.

Below we explore a route to state preparation and stabilization, which builds upon the recently developed ``active steering'' protocols \cite{Sivak2022,Liu2022a,Friedman2023,Herasymenko2023,Ravindranath2023,Morales2024,Hauser2024}.   
(We here follow Ref.~\cite{Roy2020} and use the notion  ``steering'' as proxy for ``guiding'' the
quantum system; this notion differs from ``quantum steering'' in quantum information theory \cite{Uola2020}.) 
In active steering protocols, one employs feedback operations based on \emph{weak measurements} of the system, instead of performing projective syndrome measurements as in standard QEC. 
 Typically, in weak measurements, the system qubits of interest are weakly coupled to detector qubits which are then projectively measured   \cite{Wiseman2010}.  
The key idea is then to perform periodically repeated, non-invasive measurements of the system state, 
where the detector output determines the application of subsequent operations in each time step, e.g., in order to prepare, stabilize, or manipulate a desired quantum state. 
In the broader context of quantum control, active steering protocols are examples of state-based feedback protocols and measurement-based feedback control. This type of quantum control has a long history in quantum information and quantum optics \cite{Wiseman2010}, with one of the main applications being continuous-time error correction, see, e.g., Ref.~\cite{Borah2022} and references therein. Active steering protocols should be contrasted with ``passive steering'' protocols \cite{Roy2020,Edd2023}, where the weak measurement outcomes are simply discarded.  In effect, passive steering protocols are equivalent to the driven-dissipative passive protection schemes mentioned above. 

While active steering protocols require one to perform many measurements, which in many presently studied 
qubit platforms is a quite costly and time-consuming task \cite{Blais2021}, they also come with important benefits.  
On the one hand, under fairly general conditions, it is impossible to target certain states, e.g., 
Green-Horne-Zeilinger (GHZ) states, by driven-dissipative or passive steering protocols \cite{Ticozzi2012},
while they are accessible to active steering \cite{Morales2024}.  We note that for stabilizer states like GHZ states, simpler state preparation schemes are available, see, e.g., Refs.~\cite{Zhu2023,Kam2024,chen2023}. The results of Ref.~\cite{Ticozzi2012} do not apply to such routes. Importantly, active steering could also work for non-stabilizer states.

In the present work, however, we focus on small (few-qubit) systems and explore
the impact of a finite error rate $\Gamma$ on active steering.
We consider amplitude and/or phase noise, but it is straightforward to add other error channels. 
The goal of our protocol is to prepare and stabilize, for arbitrarily long times, a predesignated 
pure target state.  Such protocols have previously been studied for circuits with up to $N=6$ system qubits in
the idealized error-free ($\Gamma=0$) case \cite{Morales2024}.   
A key result of our work is the demonstration of a \emph{sharp error threshold} (characterized, e.g., by a discontinuous derivative with respect to the error strength $\Gamma$), present already for just one or two
system qubits.  The threshold rate, $\Gamma=\Gamma_c$, separates a weak-damping regime ($\Gamma<\Gamma_c$)  which realizes 
a variant of autonomous state stabilization, from a strong-damping regime ($\Gamma>\Gamma_c$) 
where the target state cannot be reached anymore.
The existence of the error threshold is diagnosed using two quantities, namely the fidelity and the purity \cite{Schumacher_1996,Lloyd_1997,Fan2024} of the late-time state $\overline{\rho}$ averaged over many measurement trajectories.  For $\Gamma\ll \Gamma_c$, we obtain nearly perfect fidelity and purity, which then gradually decrease
upon increasing $\Gamma$.  At $\Gamma=\Gamma_c$, we observe a closure of the purity gap \cite{Bardyn2018,Sieberer2023} and thus have an infinite-temperature state, i.e., a maximally mixed state.    Depending on the detailed error channels, for $\Gamma>\Gamma_c$, one either approaches a pure dark state or, if only phase noise is present, one simply remains at the infinite-temperature state.

Our results suggest that active steering can implement autonomous state stabilization as long as the error rate remains sufficiently small.  
While we study a concrete realization in terms of Andreev qubits, our results are of general nature since we consider a generic model of system and detector qubits coupled by Pauli operators.  In order to provide estimates for realistic threshold error rates, we perform numerical simulations for Andreev qubits \cite{Desposito2001,Nazarov2003,Padurariu2010} coupled to a photon cavity detector (LC circuit)
\cite{Rasmussen2021,Blais2021}. Andreev qubits can be encoded by the subgap Andreev bound states (ABSs) in nanoscale superconducting point contacts, as discussed in theoretical  \cite{Alvaro2011,Zazunov2014,Olivares2014,Kurilovich2021,Park2017,Ackermann2023,Kate2024} and experimental \cite{Janvier2015,Woerkom2017,Park2020,Hays2020,Hays2021,Poschl2022,Fatemi2022,Matute2022,Wesdorp2023,Pita2023,Wesdorp2024,Cheung2023} work.
For the Andreev qubit considered below, experiments have already demonstrated coherence times of $\sim 500$~ns \cite{Janvier2015}, where detector readout times of order 10 ns are possible.  

In Sec.~\ref{sec2}, we describe the active steering protocol and model it by a stochastic
master equation (SME) \cite{Wiseman2010}.  The SME describes the dynamics of the system state
$\rho(t)$ in the presence of both weak measurements and error channels.  
In our implementation of the protocol, knowledge of the time-evolving state is needed.  For small system size $N$, 
this can be achieved by a classical calculation performed in parallel to the experiment.  
In its present formulation, the protocol is therefore not scalable to large system size $N$.  
However, the error threshold transition is clearly manifest already for $N=1$ and $N=2$ system qubits.

In Sec.~\ref{sec3}, we present numerical results obtained from the SME.  We show that for both $N=1$ and $N=2$, a sharp threshold occurs at a 
(non-universal) error rate $\Gamma_c$. The precise value of $\Gamma_c$  
depends on the balance between steering processes subject to active decision making and 
measurement- as well as error-induced stochastic relaxation processes.  
Our simulation results show that the threshold behavior is clearly visible in the state fidelity and the purity.  The threshold is characterized by a purity gap closing as a function of $\Gamma/\Delta$, where $\Delta$ is the superconducting pairing gap.

Finally, in Sec.~\ref{sec4}, we offer some conclusions and provide an outlook.  We use units with $\hbar=e=1$ below.

\section{Active steering in the presence of noise} \label{sec2}

\begin{figure}
\input{f1}
\caption{Illustration of the active steering scheme for (a) $N=1$ and (b) $N=2$ system qubits.  For $N=1$, the system (``s'', orange square) qubit is initialized at time $t=0$ in the state $|0\rangle$, and the detector (``d'', blue circle) is initialized in $|0\rangle_d$. After the
time step $\delta t$, during which the Hamiltonian $H$ in Eq.~\eqref{Ham} entangles the system and detector qubits (indicated by the
dashed line connected system and detector qubits), one performs a projective measurement of the detector in the computational basis.  Based on the measured outcome, the averaged expected changes $\langle (dO_s)_\alpha\rangle_{\rm ms}$} in Eq.~\eqref{dos} 
of a system operator $O_s$ are calculated for all possible ``steering operators'' in Eq.~\eqref{sop}, with $\alpha=1,\ldots,7$, 
by a classical computation (``control unit''). We allow for error channels acting on the system qubit during the protocol.
The observable $O_s$ is chosen such that the eigenstate for its 
maximal eigenvalue is the desired target state $|\Psi_f\rangle$.  Maximizing $\langle (dO_s)_\alpha\rangle_{\rm ms}$ over the set $\{\alpha\}$ then
dictates the steering operator (and thus $H$) selected during the subsequent time step. The detector is always
re-initialized in the state $|0\rangle_d$. This scheme is iterated until reaching the designated target state $\rho_f=|\Psi_f\rangle\langle \Psi_f|$ with high fidelity and high purity (for sufficiently small error rates).  
For the $N=2$ case in panel (b), there are no direct couplings between system qubits or between
detector qubits.  Entanglement between system qubits is generated through entanglement swapping, using Bell pair measurements of the two detector qubits \cite{Morales2024}. We again allow for local error channels.
For details, see main text.

\label{fig1}
\end{figure}

In this section, we discuss our active steering protocol for $N=1$ and $N=2$ system qubits coupled to their own detector qubits.
For a schematic illustration of the protocol, see Fig.~\ref{fig1}.
In what follows, the goal is to prepare and stabilize 
the system qubits in a predesignated pure target state $|\Psi_f\rangle$
with optimal fidelity. 
In contrast to previous work \cite{Morales2024}, we take into account error channels.  
Below, we separately discuss the cases $N=1$ and $N=2$,
see Sec.~\ref{sec2a} and \ref{sec2b}, respectively. 
As concrete application, we apply our general approach to Andreev qubit setups.
In Sec.~\ref{sec2c}, we discuss our numerical approach for solving the SME. 

\subsection{Single system qubit} \label{sec2a}

We start with the simplest case of a single system qubit with the computational basis states $\{|0\rangle,|1\rangle\}$, 
which couples to a single detector qubit with basis states $\{|0\rangle_d,|1\rangle_d\}$. (Of course, one may also use qudits instead, but for concreteness, we focus on qubits.)  We recall $\sigma^z|0\rangle=|0\rangle$ and $\sigma^z|1\rangle=-|1\rangle$, where the Pauli matrices $\sigma^{x,y,z}$  act in the system qubit space.  
Similarly, we define Pauli matrices $\tau^{x,y,z}$ in the detector qubit space.
To keep the notation simple, identity matrices are kept implicit below. 
We note in passing that already a single weakly monitored qubit reveals remarkably rich physics \cite{Gornyi2024}.

\subsubsection{General formulation}

We consider a protocol where one first initializes the system qubit in a pure state, say, $|\Psi(t=0)\rangle=|+\rangle$ with $\sigma^x|\pm\rangle=\pm |\pm\rangle$, and the detector qubit in the state $|0\rangle_d$. 
The active steering protocol for preparing and dynamically stabilizing a predesignated target state $|\Psi_f\rangle$, see also Ref.~\cite{Morales2024}, consists of a sequence of finite-time steps
of duration $\delta t$ as illustrated in Fig.~\ref{fig1}(a). Before each time step, one re-initializes the detector
qubit in the state $|0\rangle_d$.
During a given time step, the system and detector qubits evolve according to a Hamiltonian 
\begin{equation}\label{Ham}
    H=H_s+H_d+H_{sd},
\end{equation} 
where $H_s$ ($H_d$) acts only on the system (detector) qubit and $H_{sd}$ is a system-detector coupling.  
We write $H_s= {\bf s}\cdot \boldsymbol{\sigma}$ with 
 $\boldsymbol{\sigma}=(\sigma^x,\sigma^y,\sigma^z)$ and a real vector ${\bf s}$. Similarly, $H_d={\bf d}\cdot \boldsymbol{\tau}$ with $\boldsymbol{\tau}=(\tau^x,\tau^y,\tau^z)$. For $H_{sd}$, we use two-qubit Pauli operators or sums thereof.
  On top of the unitary time evolution induced by $H$ in Eq.~\eqref{Ham}, we will later on, see Eq.~\eqref{SME} below, also include error channels.  However, for the moment, we proceed by neglecting errors. 

For $H_{sd}\ne 0$, the time evolution during the time step $\delta t$ tends to increase the entanglement between the system and the 
detector qubit.  At the end of the time step, one performs a projective measurement of the detector Pauli matrix $\tau^z$, with outcome $\xi=0$ (if one measures the eigenvalue $+1$ of $\tau_z$) or $\xi=1$ (if one measures $-1$). The measurement outcome is intrinsically stochastic as dictated by the laws of quantum mechanics. 
After the measurement with outcome $\xi$, we have the system-plus-detector product state $|\Psi(t+\delta t) \rangle_\xi\otimes |\xi\rangle_d$, with the measurement-conditioned system state \cite{Nielsen2000}
\begin{equation}\label{Kraus1}
|\Psi(t+\delta t)\rangle_\xi =\frac{A_{\xi}}{\sqrt{P_{\xi}}} |\Psi(t)\rangle.
\end{equation}
The Kraus operator $A_\xi$ acts in the Hilbert space of the system qubit,
\begin{equation}\label{Kraus2}
A_{\xi}={}_d\langle \xi|e^{-i\delta t H}|0\rangle_d. 
\end{equation}
The corresponding probabilities for measuring the outcome $\xi$ are 
\begin{equation}\label{prob1}
    P_{\xi}=\langle \Psi(t)|A^{\dagger}_{\xi}A_{\xi}^{}|\Psi(t)\rangle.
\end{equation}
One then re-initializes the detector qubit in the state $|0\rangle_d$, such that the total initial state for the next time step is $|\Psi(t+\delta t)\rangle_\xi\otimes |0\rangle_d$, and iterates the scheme. 
An important part of the active steering protocol is the choice of the Hamiltonian $H$ in different time steps.

In the weak measurement limit \cite{Wiseman2010} of interest here, one can expand Eq.~\eqref{Kraus2} in $\|H\|\delta t\ll 1$. We first define an effective, and in general non-Hermitian, system Hamiltonian $H_{\xi=0,1}$ as
\begin{equation}\label{Hxixip}
    H_{\xi} = H_s \delta_{\xi,0} +{}_d\langle\xi|H_{sd}|0\rangle_d + {}_d\langle \xi|H_d|0\rangle_d.
\end{equation} 
We note that for the applications below, the last term in Eq.~\eqref{Hxixip} does not contribute. 
With Eq.~\eqref{Kraus2}, we obtain the Kraus operators in the form
\begin{equation}
    A_{\xi}=\sqrt{\delta t}\, \xi c+ (1-\xi) \left[1-i\delta t H_{0}-\frac{\delta t}{2} c^{\dagger}c \right]+ {\cal O}(\delta t^2),
\end{equation} 
where terms of order $\delta t^2$ are discarded after renormalizing the jump operator according to
\begin{equation}\label{jump1}
    c = -i \sqrt{\delta t}\, H_{1}.
\end{equation}
 Collecting all terms in Eq.~\eqref{Kraus1}, one arrives at a SME \cite{Wiseman2010,Morales2024}, where we next include error channels. For that purpose, we switch to the density matrix $\rho(t)$ of the system qubit since in general we now encounter mixed states even for individual measurement trajectories.

For the measurement-conditioned time evolution  after one time step, $d\rho_\xi=\rho_\xi(t+\delta t)-\rho$  with $\rho=\rho(t)$, we thereby obtain the canonical SME  \cite{Wiseman2010},
\begin{eqnarray}\label{SME}
d\rho_\xi&=&-i\delta t\, [H_0,\rho] + \xi \left(\frac{c\rho c^{\dagger}}{\langle c^{\dagger}c\rangle}-\rho\right) 
\\ & -&\frac{\delta t}{2}\{c^{\dagger}c-\langle c^{\dagger}c\rangle,\rho\} + \delta t \sum_{\gamma}\mathcal{D}[c_{\gamma}]\rho ,\nonumber
\end{eqnarray}
where $\{\cdot,\cdot\}$ is the anticommutator and 
the quantum expectation value of an operator $A$ is $\langle A\rangle={\rm Tr}\left(\rho A\right)$.  
The random variable $\xi$ describing the measurement outcome has the \emph{a priori} probabilities, cf.~Eq.~\eqref{prob1}, 
\begin{equation}\label{pr2}
    P_{\xi=1}=\delta t \langle c^\dagger c\rangle   , \quad P_{\xi=0}=1-P_{\xi=1}.
\end{equation}
The sum over the dissipation channels ($\gamma$) in Eq.~\eqref{SME} contains the standard Lindbladian superoperator \cite{Wiseman2010}, ${\cal D}[c] \rho = c \rho c^\dagger - \frac12 \{ c^\dagger c,\rho\}.$
For concreteness, we  include only two error channels below. However, it is straightforward to add other channels. First, we allow for \emph{amplitude damping} of the system qubit with rate $\Gamma_{\rm AD}$. This is described by  the jump operator $c_{\rm AD}=\sqrt{\Gamma_{\rm AD}}\,\sigma^-$ with $\sigma^-=(\sigma^x-i\sigma^y)/2$ \cite{Nielsen2000}. Second, we include \emph{dephasing} of the system qubit with rate $\Gamma_{\rm PD}$, 
where the jump operator is $c_{\rm PD}=\sqrt{\frac{\Gamma_{\rm PD}}{2}}\,\sigma^z$ \cite{Nielsen2000}.  

We note that Eq.~\eqref{SME} tacitly assumes that the measurement time required for performing projective detector measurements is short 
compared to $\delta t$.  This assumption may or may not be valid in practice, depending on the platform. Below we give concrete numbers for the case of an Andreev qubit. However,  one could 
refine our analysis by including an extra time interval after each time step, where the measurement dynamics is explicitly considered and the system qubit remains subject to decoherence and dephasing.  We leave such generalizations to future work. In addition, we neglect the reset time for the detector.

Our active steering protocol aims at maximizing a suitable system observable $O_s$ such that the target state is approached over the course of the protocol.
For instance, if we wish to prepare the state $|\Psi_f\rangle=|0\rangle$,  the 
observable is chosen as $O_s=\sigma^z$ since the maximal expectation value $\langle O_s\rangle=+1$ is only achieved for the desired target state. While we only assume pure initial states throughout the paper, it would be straightforward to extend the discussion to an arbitrary initial state, including mixed states. 
Taking a fixed system-detector coupling $H_{sd}$,  
we assume that, depending on the previous measurement record, one can select one out of seven \emph{steering operators} $H_{s,\alpha}$ (with $\alpha=1,\ldots,7$) acting on the system qubit during the following time step.  
In general, for a given time step, the system Hamiltonian $H_s$ then contains a piece $H_s^{(0)}$, which is always present, and a steering operator $H_{s,\alpha}$ (when choosing the system Hamiltonian for the next step as done here, rather than the system-detector coupling, the active decision making sometimes is referred to as  ``active shaking'' rather than ``active steering'', cf.~Ref.~\cite{Roy2020}, where shake-and-steer protocols are discussed),
\begin{equation}\label{sop}
    H_s=H_s^{(0)}+H_{s,\alpha},\quad H_{s,\alpha} \in \{ 0, \pm J\sigma^x, \pm J\sigma^y, \pm J \sigma^z \}.
\end{equation}
We allow for $H_s^{(0)}\ne 0$ since this term is 
naturally present in the Andreev qubit application studied below.
For the weak measurement limit,  the steering operator strength $J$ should satisfy $J\delta t\ll 1$.  
For simplicity, we assume a fixed constant value for $J$. 

We next describe the active decision making part of the protocol, i.e., how  $H_{s,\alpha}$ is chosen for the subsequent time step.
After a given time step has been finished with measurement outcome $\xi$, for every value of $\alpha$ in Eq.~\eqref{sop}, 
we compute the measurement-averaged change of the expectation value of the observable $O_s$ anticipated after the following time step,
\begin{equation}\label{dos}
\langle (dO_s)_\alpha \rangle_{\rm ms} =\sum_{\xi=0,1} P^{(\alpha)}_{\xi} \mathrm{Tr}\left(d\rho^{(\alpha)}_\xi O_s\right), 
\end{equation}
where $\langle O\rangle_{\rm ms}$ indicates an average over measurement outcomes.
Here, $d\rho_\xi^{(\alpha)}$ follows from Eq.~\eqref{SME} with $H_s=H_s^{(0)}+H_{s,\alpha}$, and the probabilities $P_\xi^{(\alpha)}$ are given in Eq.~\eqref{pr2}.  For the next time step,
we then select the steering operator (indexed by $\alpha$) which maximizes Eq.~\eqref{dos}. In this
manner, on average, we effectively also maximize $\langle O_s\rangle$ during the course of the protocol.

The above scheme allows for error channels but requires a 
classical calculation performed in parallel to the experiment after every time step.
As shown by numerical simulations in Sec.~\ref{sec3}, one approaches the target state (which is the eigenstate for the maximum eigenvalue of $O_s$) with high fidelity unless error rates become too large. 
 
\begin{figure}
\includegraphics[scale=0.9]{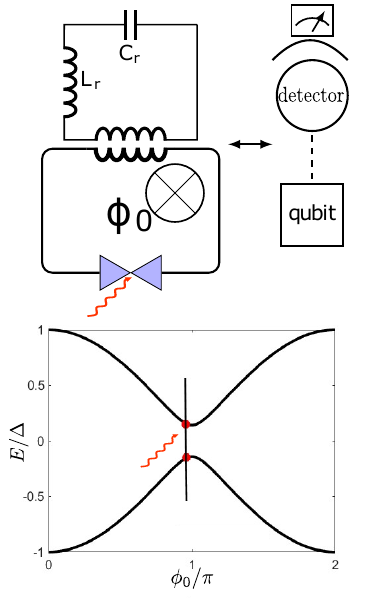}
\caption{Example for the $N=1$ protocol: The two ABS levels of a phase-biased short nanoscale superconducting junction harbor an Andreev qubit (system qubit), which is 
coupled to the electromagnetic phase fluctuations of an LC circuit. In a parameter regime allowing for the truncation to only $\xi=0,1$ photons in the resonator, the LC circuit serves as detector qubit. 
Upper panel: The left side shows a sketch of the experimental setup, where the magnetic flux threading a superconducting loop containing the junction imposes the average phase difference $\phi_0$ across the junction. 
The LC circuit is inductively coupled to this loop.  The right side corresponds to the equivalent schematic 
circuit shown in Fig.~\ref{fig1}(a). Red arrows indicate the steering operator $H_{s,\alpha}$, see Eq.~\eqref{sop}, applied to the point contact which may induce transitions between system qubit states. 
Lower panel: ABS dispersion relation $E=\pm E_A(\phi_0)$ (in units of the superconducting gap $\Delta$) vs 
superconducting phase difference $\phi_0$ for transmission probability ${\cal T}=0.98$, see Eq.~\eqref{ABS}. For the implementation of the steering protocol, we fix $\phi_0=0.97\pi$ as indicated by red dots. }
\label{fig2}
\end{figure}

\subsubsection{Application: Andreev qubit}\label{andreevsec}

As concrete example for our general approach, we consider an Andreev qubit 
coupled to an LC resonator operated in the single photon limit, thus acting as an effective detector qubit. 
The Andreev qubit is realized from the ABS pair in  a short nanoscale superconducting point contact \cite{Desposito2001,Nazarov2003,Padurariu2010,Zazunov2014}.
The schematic measurement setup and the ABS dispersion as function of the superconducting phase difference $\phi_0$ are shown in Fig.~\ref{fig2}.  For experimental realizations, see Refs.~\cite{Janvier2015,Woerkom2017,Park2020,Hays2020,Hays2021,Poschl2022,Fatemi2022,Matute2022,Wesdorp2023,Pita2023,Wesdorp2024}.
In eigen-energy representation, the system Hamiltonian (without steering operators) is written as 
\begin{equation}\label{hs0}
    H_s^{(0)}=E_A(\phi_0)\, \sigma^z,
\end{equation}
with the positive ABS energy \cite{Beenakker1991,Furusaki1991,Nazarov2009,Alvaro2011} 
\begin{equation}\label{ABS}
    E_A( \phi_0) =\Delta \sqrt{1-{\cal T}\sin^2(\phi_0/2)},
\end{equation}
where $\Delta$ is the superconducting gap and ${\cal T}$ the 
transmission probability of the contact. The particle-hole partner state is also shown in the lower panel of
Fig.~\ref{fig2}.
The Andreev qubit is defined by the ground and excited states according to $|g\rangle=|1\rangle$ and $|e\rangle=|0\rangle$, respectively. 

The detector qubit is modeled by an LC circuit with the Hamiltonian $H_d=\Omega a^{\dagger}a$, where
the photon number is constrained to $a^{\dagger}a\in\{0,1\}$, i.e., we assume that parameters are chosen 
such that higher photon numbers do not occur with significant weight.  Although this condition may be difficult to meet in actual physical implementations of microwave resonators coupled to Andreev qubits \cite{Janvier2015}, we note that our model also describes the coupling of the Andreev qubit to a transmon detector qubit as used in recent experiments \cite{Pita2023}.
In any case, one can equivalently write $H_d$ (up to an irrelevant energy shift) in qubit language as $H_d=-\Omega \tau^z$, where $|0\rangle_d$ ($|1\rangle_d$) is the state with $\xi=0$ ($\xi=1$) photons.    The
frequency $\Omega$ does not influence our protocol since $H_d$ only generates an irrelevant global phase.

Finally, the system and detector qubits are weakly coupled by  
\begin{equation}\label{hsd1}
    H_{sd}= \lambda_0 \hat{I}_s \tau^x = \Lambda \, \sigma^s \tau^x,
\end{equation}
with a coupling constant $\lambda_0$ and the ABS supercurrent operator $\hat I_s=I_0 \sigma^s$. Here,
$\sigma^s$ is a rotated Pauli matrix and $I_0$ denotes the supercurrent scale \cite{Zazunov2014},
\begin{eqnarray}\nonumber
\sigma^s &=& \frac{\Delta}{E_A(\phi_0)}\left[-\cos(\phi_0/2)\sigma^z+\sqrt{1-{\cal T}}\sin(\phi_0/2)\sigma^y\right],\\ 
 I_0&=& {\cal T}\Delta \sin(\phi_0/2).\label{curop}
\end{eqnarray} 
In Eq.~\eqref{hsd1}, we have introduced the energy scale $\Lambda=\lambda_0 I_0$ for the system-detector coupling.
In the weak measurement limit, $\Lambda \delta t\ll 1$.
The steering operators $H_{s,\alpha}$ in Eq.~\eqref{sop} could be realized in experiments by means of flux or gate driving pulses along the lines discussed in Ref.~\cite{Metzger2021}.

For the results in Sec.~\ref{sec3}, we assume a transmission probability ${\cal T}=0.98$ and a
phase difference $\phi_0=0.97\pi$. 
For $\phi_0\sim\pi$ and ${\cal T}$ close to unity, the Andreev qubit energies $\pm E_A$ in Eq.~\eqref{ABS} are very small.
Since $E_A\ll \Delta$, detrimental transition rates involving above-gap continuum quasiparticles are strongly suppressed  and therefore long qubit coherence times are expected \cite{Zazunov2003,Zazunov2014,Ackermann2023}. 
Instead of simply putting $\phi_0=\pi$ and ${\cal T}=1$, we consider a generic choice,
where the system-detector coupling \eqref{hsd1} contains $\sigma_y$ and $\sigma_z$.

On the other hand, we assume negligibly short detector readout and detector reset time scales below. (In case one wants to include the effects of those times explicitly,
one has to add the uncoupled but noisy  evolution of the system qubits during the measurement and reset periods in the SME.) 
While this assumption is rather unrealistic for transmon qubits, where a readout time $\sim 100$~ns currently sets the optimal value \cite{Rasmussen2021}, great progress has been achieved for fast ($<10$~ns) dispersive readout of semiconducting quantum dot qubits \cite{Horstig2024}.  Detector reset times are typically much faster \cite{Magnard2018}.
We note that the time step $\delta t$ of the protocol should be chosen
large compared to the measurement (and reset) time of the detector. In fact, the dimensionless parameter $\Delta \delta t$ is limited from both below and from above.   The lower limit arises from the measurement and reset) time and, in numerical computations as presented below, from the fact that otherwise the numerical run-times necessary for converging the protocol towards the steady state become very long.  The upper limit instead is imposed by the weak measurement limit assumed in our formulation of the theory.
For these reasons, we explore only a relatively narrow window of values for $\Delta \delta t$ in Sec.~\ref{sec3} below.  However, already for those values, we find indications for ``universal''
error thresholds.    

\subsection{SME for two system qubits}\label{sec2b}

In Sec.~\ref{sec3}, we also show simulation results for $N=2$ system qubits. As shown in Fig.~\ref{fig1}(b), each system qubit is here only coupled to its own detector qubit, without direct couplings neither between the two system qubits nor between the two detector qubits.  The Hamiltonian for $N=2$ can thus be written as
$H=\sum_{n=1,2} H^{(n)}$ with $H^{(n)}$ as in Eq.~\eqref{Ham}. 
Below, $\boldsymbol{\sigma}_n$ refers to the Pauli matrices for system qubit $n=1,2$, and similarly $\boldsymbol{\tau}_n$ for 
detector qubits.  The protocol starts with a pure initial state of the system qubits, say, $|\Psi(0)\rangle=|++\rangle$, and
with the state $|00\rangle_d$ for the detector qubits.  Before each time step, the detector qubits are re-initialized in this state again.

We here discuss the case where weak measurements of the system
qubits are performed by projective Bell pair measurements of the two detector qubits,
see Ref.~\cite{Morales2024} for a detailed discussion of this scheme.
Entanglement between the two system qubits is generated by a weak-measurement variant of entanglement swapping \cite{Horodecki2009,Huang2020,Morales2024}, where the detector qubits are projected onto one 
of the four maximally entangled Bell states, 
\begin{eqnarray}\nonumber
    |\Phi_{\xi=0,\eta=\pm}\rangle_d &=& (|00\rangle_d\pm |11\rangle_d)/\sqrt{2}, \\
    |\Phi_{\xi=1,\eta=\pm}\rangle_d &=& (|01\rangle_d\pm |10\rangle_d)/\sqrt{2}, \label{Bell}
\end{eqnarray} 
at the end of each time step.  Measurement outcomes are thus characterized by the two numbers
$\xi\in \{0,1\}$ and $\eta\in\{-1,+1\}$.
The four Bell states \eqref{Bell} have either even ($\xi=0$) or odd ($\xi=1$) parity, and are either symmetric ($\eta=+1$)
or antisymmetric ($\eta=-1$) under an exchange of both qubits. (Here, even parity means that states are built from the 
basis states $|00\rangle_d$ and $|11\rangle_d$, and likewise for odd parity.)  Such Bell measurements can be implemented by 
measuring the two commuting Pauli two-qubit operators 
\begin{equation}\label{BellOP}
    \hat {\cal O}^x=\tau_1^x\tau_2^x,\quad \hat {\cal O}^z=\tau^z_1\tau^z_2.
\end{equation}
Indeed, one readily finds 
\begin{equation}
    \hat {\cal O}^x|\Phi_{\xi,\eta}\rangle_d
=\eta|\Phi_{\xi,\eta}\rangle_d,\quad \hat {\cal O}^z|\Phi_{\xi,\eta}\rangle_d
=(1-2\xi) |\Phi_{\xi,\eta}\rangle_d.
\end{equation} 
For the Andreev qubit platform with cavity detectors, the
implementation of the above Bell measurements could be performed along the lines of Ref.~\cite{Riste2013}.

One can then construct the SME for the state $\rho(t)$ of the two system qubits in close analogy to Sec.~\ref{sec2a}, 
see Eq.~\eqref{SME}.  In analogy to Eq.~\eqref{Hxixip}, we define the effective system Hamiltonian
$H_{\rm eff}=\sum_{n=1,2} H_0^{(n)}$ with  
\begin{equation}\label{hxi2}
    H_{\xi}^{(n)}  = \sqrt2 \,\,  {}_d\langle \Phi_{\xi,\eta} |H^{(n)}|00\rangle_d, 
\end{equation} 
 for the $n$th system qubit. Here Eq.~\eqref{hxi2} is independent of $\eta$ except for the case $n=1$ and $\xi=1$, where we
can write $\eta H_{1}^{(1)}$ for the matrix element. The SME for $d\rho_{\xi,\eta} = \rho_{\xi,\eta}(t+\delta t)-\rho$ conditioned on the measurement outcome follows in almost identical form as in Eq.~\eqref{SME},
\begin{eqnarray}\label{SME2}
d\rho_{\xi,\eta}&=&-i\delta t\, [H_{\rm eff},\rho] +
\xi \left(\frac{c_\eta \rho c_\eta^{\dagger}}{\langle c_\eta^{\dagger}c_\eta^{}\rangle}-\rho\right) 
\\ & -&\frac{\delta t}{2}\{c_\eta^{\dagger}c_\eta^{}-\langle c_\eta^{\dagger}c_\eta^{}\rangle,\rho\} + 
\delta t \sum_{\gamma}\mathcal{D}[c_{\gamma}]\rho .\nonumber
\end{eqnarray}
The jump operators $c_\eta^{}$ for outcome $\eta=\pm$ are given by
\begin{equation}\label{nonunit}
    c_{\eta}=-i\sqrt{\delta t}\left(\eta H^{(1)}_1 +  H^{(2)}_1\right).
\end{equation}
In analogy to Eq.~\eqref{prob1}, the probability for measuring $(\xi,\eta)$ follows as
\begin{equation}
 P_{\xi,\eta}=\frac12\left(\delta_{\xi,0}+(\delta_{\xi,1}-\delta_{\xi,0})\delta t\langle c_\eta^\dagger c_\eta^{}\rangle\right).
 \end{equation}
While the unitary jump operator \eqref{jump1} in the single-qubit case considered here does not allow for state purification, the non-unitary jump operators for the two-qubit case in Eq.~\eqref{nonunit} admit this possibility. 
In fact, for unitary jump operators $c=U$ with $U^\dagger U=1$ describing measurement effects in the SME, purity is preserved since ${\rm Tr}\rho^2={\rm Tr}(U\rho U^\dagger)^2$.  This argument holds as long as one neglects amplitude damping and the interplay with the Hamiltonian dynamics.  However, if a quantum jump happens, the dynamics is dominated by 
the corresponding contribution $\propto \xi$ in Eq.~\eqref{SME2}.
We note that while the single-qubit jump operator \eqref{jump1} can in general also be non-unitary, 
for the Andreev qubit application, $(\sigma^s)^2=1$ in Eq.~\eqref{curop} implies unitarity.  The two-qubit case, in contrast, generically comes with non-unitary jump operators.
Accordingly, for weak error rates, we find that the active steering protocol generally performs better for $N=2$ than for $N=1$ because of the purifying capabilities of the non-unitary $N=2$ jump operators. 

For the SME  \eqref{SME2}, we again introduced dissipation terms such as dephasing and amplitude damping.
We assume that those channels separately act on each system qubit $\boldsymbol{\sigma}_{n=1,2}$. 
The jump operators for dephasing are given by $c_{\rm PD}^{(n)}=\sqrt{\frac{\Gamma_{\rm PD}^{(n)}}{2}}\,  \sigma_n^z$, 
with the corresponding rate $\Gamma_{\rm PD}^{(n)}$. Similarly, for amplitude damping, we have 
$c_{\rm AD}^{(n)}=\sqrt{\Gamma_{\rm AD}^{(n)}}\, \sigma_n^-$.  The summation over $\gamma$ in Eq.~\eqref{SME2} includes
both the sum over $n=1,2$ and over those two error channels.   
For simplicity, we assume below that the noise strength is identical for each qubit, i.e.,
$\Gamma_{\rm AD}^{(n=1,2)}=\Gamma_{\rm AD}$ and $\Gamma_{\rm PD}^{(n=1,2)}=\Gamma_{\rm PD}$.

For the active decision making, we proceed as in Sec.~\ref{sec2a} and 
maximize the measurement-averaged (anticipated) change 
of the expectation value of a suitable system observable $O_s$ after the next time step, cf.~Eq.~\eqref{dos}, 
\begin{equation}
 \langle (dO_s)_\alpha\rangle_{\rm ms} =\sum_{\xi,\eta} P_{\xi,\eta}^{(\alpha)}\, {\rm Tr} \left( d\rho^{(\alpha)}_{\xi,\eta} \, O_s \right) 
\end{equation}
over the set $\{\alpha\}$ of possible steering operators.  In order to target a Bell state, see Eq.~\eqref{Bell} but now for 
system qubits, we employ one of the four operators, cf.~Eq.~\eqref{BellOP}, 
\begin{equation}
    O_s=\pm \sigma_1^x \sigma_2^x \pm \sigma_1^z \sigma_2^z.
\end{equation}
For instance, by choosing both $+$ signs and maximizing $O_s$, one  
targets the symmetric even-parity Bell state 
\begin{equation}\label{bell2}
    |\Psi_f\rangle=|\Phi_{0,+}\rangle=(|00\rangle+|11\rangle)/\sqrt2. 
\end{equation}
We here consider the experimentally simplest case, where steering operators only act independently on individual system qubits,
cf.~Eq.~\eqref{sop} and Fig.~\ref{fig1}.  We thus assume system Hamiltonians of the form $H_s = \sum_{n=1,2} \left( H_{s,n}^{(0)} + H_{s,\alpha_n} \right)$, where
$H_{s,n}^{(0)}$ is given by Eq.~\eqref{hs0} with $\sigma^z\to \sigma_n^z$
and the index $\alpha=(\alpha_1,\alpha_2)$ now covers $7^2=49$ possibilities,
cf.~Eq.~\eqref{sop}.\\ 
 
\subsection{Numerical solution of the SME}\label{sec2c}

In Sec.~\ref{sec3}, we present numerical simulation results based on the active steering protocol   for   $N=1$, see Sec.~\ref{sec2a}, and for $N=2$, see Sec.~\ref{sec2b}. 
Using the SME in Eq.~\eqref{SME}, or the one in Eq.~\eqref{SME2}, in the presence of error channels, the measurement-resolved trajectory is described by a mixed state $\rho(t)$
at times $t=j\delta t$  (integer $j>0$).  
For an initially pure state $|\Psi(0)\rangle$, however, one can unravel the 
SME into an equivalent stochastic Schr\"odinger equation (SSE) \cite{Breuer2000,Semin2017,Gisin_1992,Stefano2008,Campaioli2024} 
by resolving pure state trajectories $|\Psi(t)\rangle$ for the measurement dynamics and for the environmental (``bath'') stochastic forces which generate the dissipative terms in the SME.  In numerical implementations of the SME, it is advantageous to employ the SSE formulation \cite{Morales2024,Landi2024,Campaioli2024}.  
 By averaging the dynamics predicted by the SSE over the stochastic bath variables, one recovers the SME.  

We here specify the SSE only
for $N=1$, since the case $N=2$ follows analogously.  Writing
$|d\Psi\rangle_{\xi,x_\gamma}=|\Psi(t+\delta t)\rangle_{\xi,x_\gamma}-|\Psi\rangle$ with $|\Psi\rangle=|\Psi(t)\rangle$, one finds \cite{Breuer2000,Semin2017,Gisin_1992,Stefano2008,Campaioli2024}
\begin{widetext}
\begin{eqnarray}\label{SMEunravel}
|d\Psi\rangle_{\xi,x_\gamma} &=&\Biggl\{-i\delta t \,H_{0} -\frac{\delta t}{2}
\left(c^{\dagger}c-\langle c^{\dagger}c\rangle\right)+
\xi\left(\frac{c}{\sqrt{\langle c^{\dagger}c\rangle}}-1\right) + \\ 
\nonumber &+&\sum_{\gamma} \left[ \delta t\left( 2\langle c^{\dagger}_{\gamma}\rangle 
c^{}_{\gamma}-c^{\dagger}_{\gamma} c^{}_{\gamma}-\langle c^{\dagger}_{\gamma}\rangle 
\langle c_{\gamma}\rangle \right) +\sqrt{\delta t}\, x_{\gamma}\left(c_{\gamma}-
\langle c_{\gamma}\rangle\right)\right]\Biggr\}|\Psi\rangle,
\end{eqnarray}
\end{widetext}
where stochastic bath forces are represented by the normal-distributed random variables $x_{\gamma} \in \mathcal{N}(0,1)$. These variables are independent of the measurement outcome $\xi$. Indeed, our numerical analysis confirmed explicitly 
that $\xi$ and $\{x_{\gamma}\}$ remain uncorrelated. In contrast to the SME, the SSE~\eqref{SMEunravel}
 only evolves pure states and is numerically stable. 
 
 An individual measurement-resolved trajectory $\rho(t)$ solving the SME is thereby expressed in terms of an average over $n_w\gg 1$ pure-state trajectories $|\Psi(t)\rangle$ solving the SSE \eqref{SMEunravel} for different realizations of the stochastic bath variables $x_\gamma(t)$ (but the same measurement outcomes).  The average over $\{ x_\gamma(t) \}$ mimics the effect of the corresponding dissipative contributions in the SME.  
 In Sec.~\ref{sec3}, we choose $n_w=100$. This value is sufficiently large to 
 ensure convergence to the limit described by the SME without slowing down the numerical simulation too much. We have checked that this value also suffices to describe the largest damping rates 
 discussed in Sec.~\ref{sec3}. For a detailed discussion of the convergence properties with respect to $n_w$, see
 Ref.~\cite{Campaioli2024}.
 On top of the bath average, we may also average over 
 measurement trajectories in order to compute
 the averaged state $\overline{\rho(t)}$ and/or averaged physical quantities.

\section{Error threshold} \label{sec3}

In this section, we present numerical simulation results for the active steering protocols  in Sec.~\ref{sec2}. In Sec.~\ref{sec3a}, we study the case of a single Andreev qubit, see Sec.~\ref{andreevsec}, followed by the two-qubit case in Sec.~\ref{sec3b}.  For clarity, for $N=2$, we assume identical parameters for both qubit-detector subsystems and also identical  steering couplings ($J_1=J_2=J$).  In general, we find that our protocol is most efficient and robust if 
$J$ and the system-detector coupling strength $\Lambda$ are of comparable magnitude, not differing by more than one order of magnitude.  Within these bounds, the detailed choice of these parameters is not important, ensuring the robustness of the protocol.  One can rationalize the robustness of the scheme for $J\approx \Lambda$ by noting that the unitary active feedback operations will be inefficient for small $J$ while measurement-based feedback operations become inefficient for small $\Lambda$.  A good compromise is reached for $J\approx \Lambda$.
Below, we either consider equal error rates for amplitude and phase noise, $\Gamma_{\rm AD}=\Gamma_{\rm PD}=\Gamma$,
or we study the case of only phase noise with $\Gamma_{\rm AD}=0$.  
 
\subsection{One system qubit}\label{sec3a}

\begin{figure}
\includegraphics[width=\linewidth]{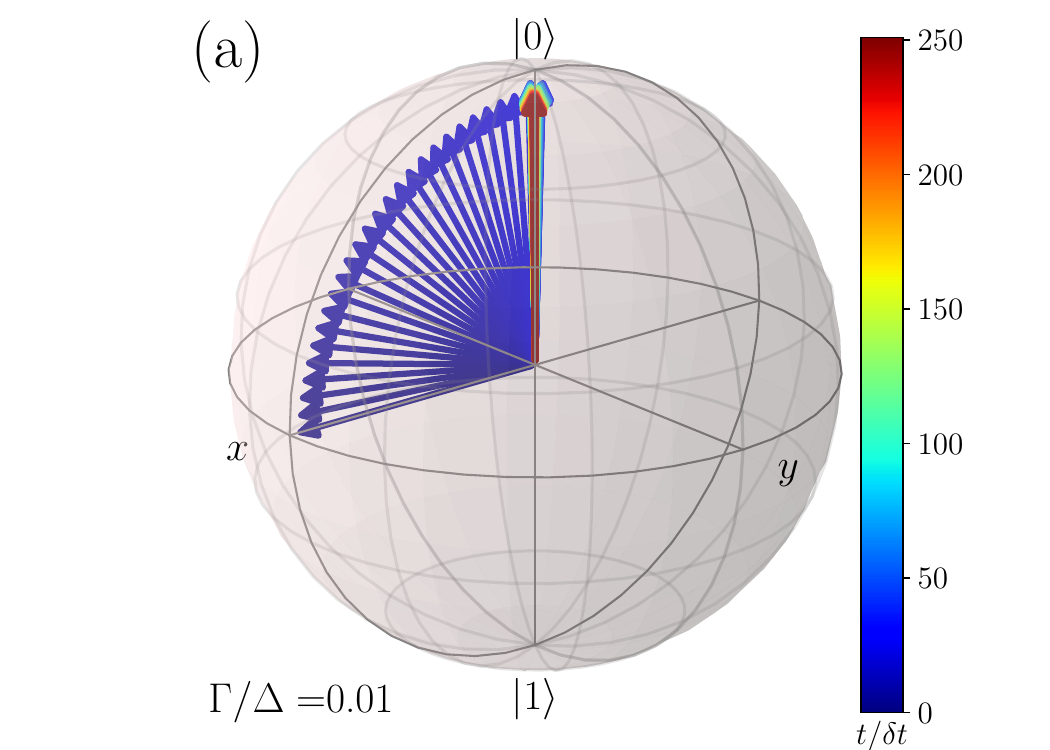}
\includegraphics[width=\linewidth]{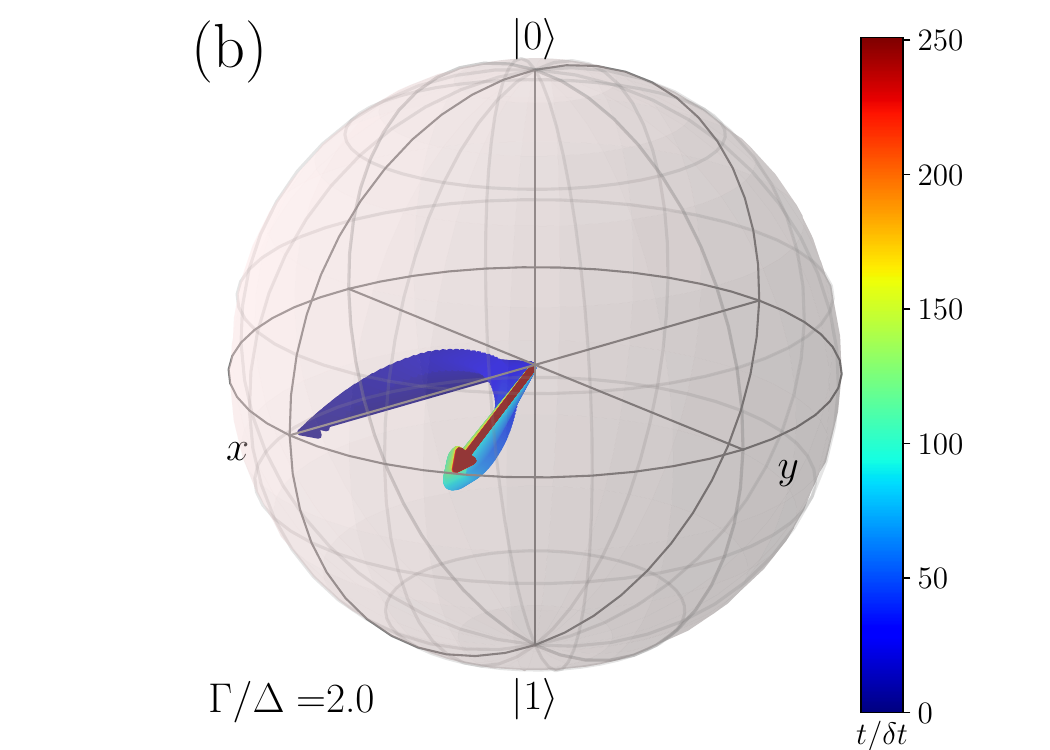}
\includegraphics[width=\linewidth]{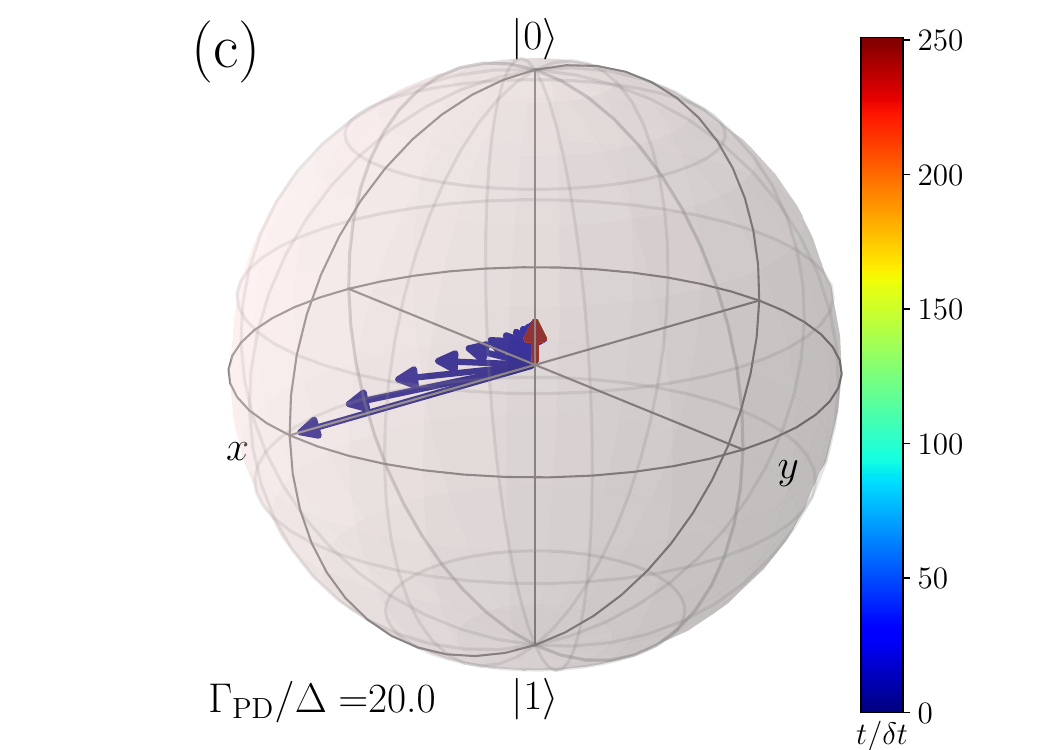}
\caption{Time evolution of the system qubit state $\overline{\rho(t)}$ averaged over $1000$ measurement trajectories of the $N=1$ active steering protocol in the presence of error channels.  Starting from $|\Psi(0)\rangle=|+\rangle$, the target state is $|\Psi_f\rangle=|0\rangle$.  We show the time evolution of $\overline{\rho(t)}$ in the Bloch unit ball for ${\cal T}=0.98, \phi_0=0.97\pi, \Lambda/\Delta=0.98, J/\Delta= 3, \Delta \delta t=0.01$,
and three choices for the error rates.
Color scales correspond to the number of time steps, see the respective color bars.
(a) Both amplitude damping and dephasing are present, $\Gamma_{\rm AD}=\Gamma_{\rm PD}=\Gamma$, with $\Gamma/\Delta=0.01$. (b) Same as panel (a) but for $\Gamma/\Delta=2$. (c) Only dephasing is present, $\Gamma_{\rm AD}=0$, with $\Gamma_{\rm PD}/\Delta=20$.   }
\label{fig3}
\end{figure}

Representative numerical simulation results for the time evolution are illustrated in Fig.~\ref{fig3}, where we 
show the system state $\overline{\rho(t)}$ averaged over $1000$ measurement trajectories. 
Using the Bloch representation, we write $\overline{\rho(t)}=\frac12( \sigma^0+ {\bf r(t)}\cdot \boldsymbol{\sigma})$ with the identity $\sigma^0$. In Fig.~\ref{fig3}, we track the vector ${\bf r}(t)$ in the Bloch unit ball, $|{\bf r}(t)|\le 1$, at different time steps. For all panels in Fig.~\ref{fig3}, we start from 
$|\Psi(0)\rangle=|+\rangle$ and consider the target state $|\Psi_f\rangle=|0\rangle$, corresponding to  $O_s=\sigma_z$ for the active decision making scheme in Sec.~\ref{sec2a}. Note that this target state represents an excited state of the Andreev qubit Hamiltonian \eqref{hs0}. In fact, we find that the active steering protocol approaches an arbitrary predesignated target state with similar convergence rate. 

We first study what happens if both amplitude damping and dephasing are present, with $\Gamma_{\rm AD}=\Gamma_{\rm PD}=\Gamma$. For weak damping,  see Fig.~\ref{fig3}(a), the active steering protocol is able to converge the averaged state towards $|\Psi_f\rangle$ with high fidelity, despite of the error channels.  
We note that $\overline{\rho(t)}$ always remains approximately pure since the Bloch vector stays near the surface of the Bloch ball. 

On the other hand, for strong damping, see 
Fig.~\ref{fig3}(b),  $\overline{\rho(t)}$ does \emph{not} converge anymore to the desired target state.  
We observe that ${\bf r}(t)$ first leaves the Bloch sphere and enters the interior of the Bloch ball, corresponding to a mixed state. At some time $t$, an infinite-temperature state with ${\bf r}(t)=0$ is reached. 
However, the final state $\overline{\rho(\infty)}$ realized at long times is (almost) pure again. 
This state corresponds to a dark state  \cite{Zanardi2014} for the SME in Eq.~\eqref{SME}.  (A dark state is a pure state which is a common zero mode of the entire set of Lindblad operators and an eigenstate of the Hamiltonian.)
The long-time dynamics of $\overline{\rho(t)}$ is thus dominated by error channels, and 
the final state reached by the protocol is basically unrelated to $|\Psi_f\rangle$.  For $\Gamma\to \infty$, one would approach the state $|1\rangle$, which is annihilated by the jump operator $c_{\rm AD}\propto \sigma^-$ and also is an eigenstate of $c_{\rm PD}\propto \sigma_z$.  However, for the parameters in Fig.~\ref{fig3}(b), measurement-induced processes are still important and yield a different dark state.
In any case, we observe qualitatively distinct regimes for the dynamics of $\overline{\rho(t)}$
for weak and strong error rates.

Finally, in the absence of amplitude damping, strong phase damping steers the system to a featureless infinite-temperature state, see Fig.~\ref{fig3}(c). In our case, the Bloch vector ${\bf r}(\infty)$ does not precisely 
vanish because the rate $\Gamma_{{\rm PD}}$ is not extremely large, and thus
measurement-induced processes are still significant.
Since the dark space of the Lindbladian for phase damping consists of the $z$-axis 
inside the Bloch ball, purity is conserved once this space has been reached.  For this reason,
${\bf r}(\infty)$ is parallel to the $z$-axis in Fig.~\ref{fig3}(c).

\begin{figure}
\includegraphics[scale=0.5]{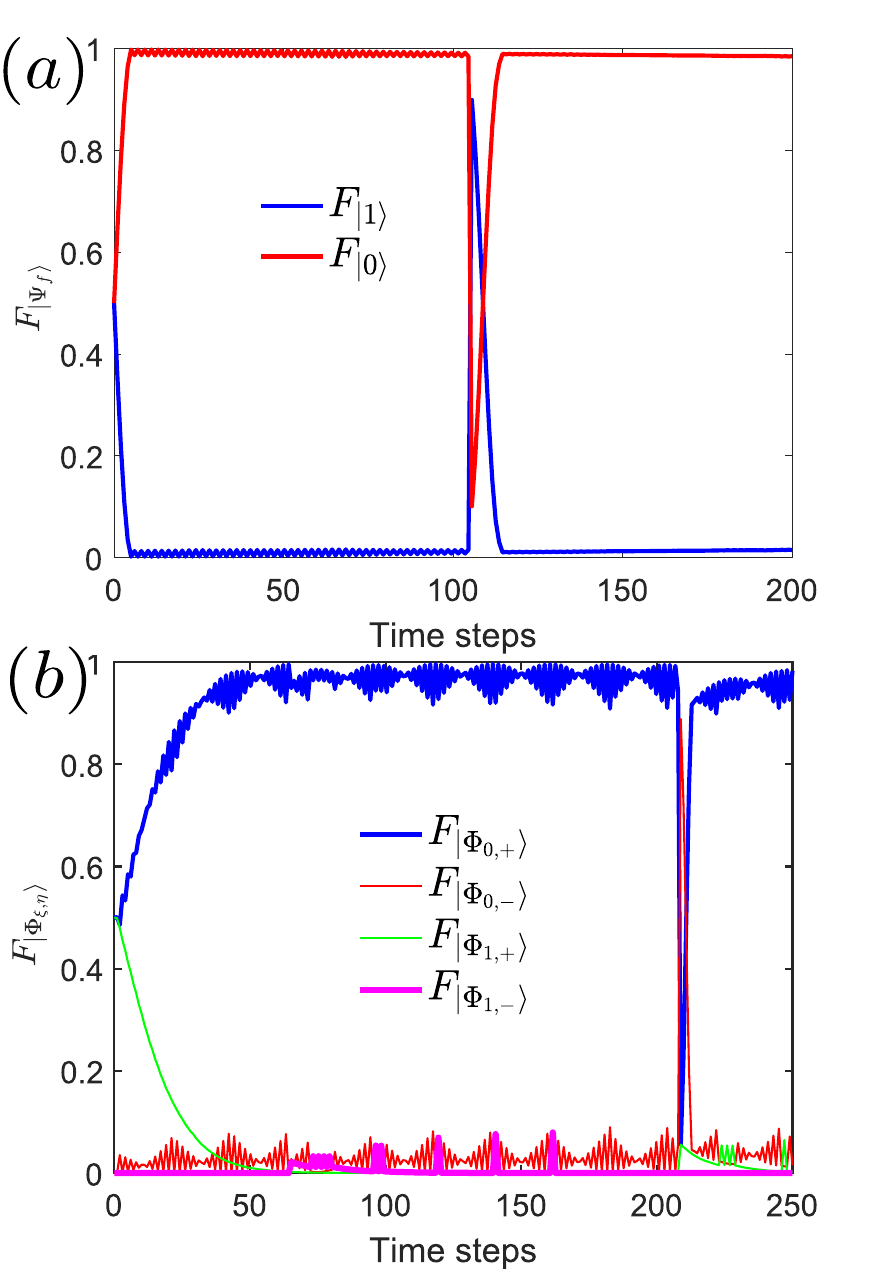}
\caption{Time-dependent fidelities, see Eq.~\eqref{fidelity}, obtained from an individual measurement trajectory for $N=1$ [panel (a)] and $N=2$ [panel (b)] system qubits under the active steering protocol with error rate $\Gamma/\Delta=  0.001$, where $\Gamma_{\rm AD}=\Gamma_{\rm PD}=\Gamma$. 
We use the same parameters as in Fig.~\ref{fig3} but with $\Delta \delta t=0.05$. 
Panel (a):  Fidelity vs number of times steps for $N=1$ with respect to the states $|0\rangle$ and $|1\rangle$, with target state $|\Psi_f\rangle=|0\rangle$.
Panel (b):  Same as in panel (a) but  for $N=2$. Both system-plus-detector subsystems in Fig.~\ref{fig1} have identical couplings $\Lambda/\Delta=0.49$. The initial system state is $|\Psi(0)\rangle=|++\rangle$ and the target state is the symmetric even-parity Bell state $|\Phi_{0,+}\rangle$ in Eq.~\eqref{bell2}.  We show the fidelities with respect to each of the four Bell states. }
\label{fig4}
\end{figure}

\begin{figure*} 
\includegraphics[scale=0.42]{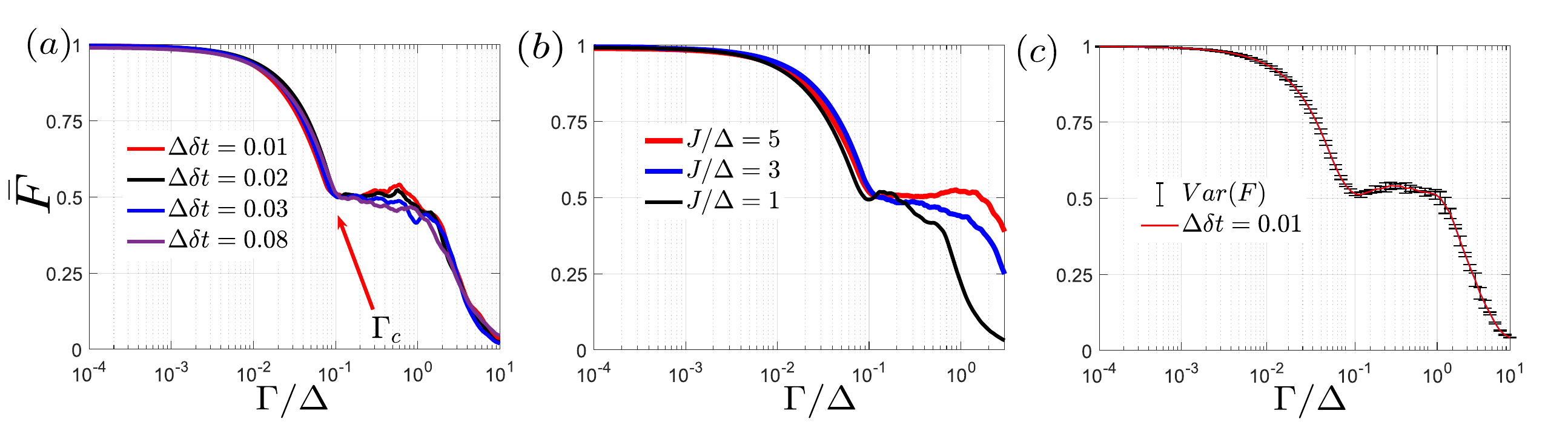}
\caption{Error rate dependence of the fidelity reached at late times of  the $N=1$ steering protocol with $|\Psi(0)\rangle=|+\rangle$ and $|\Psi_f\rangle=|0\rangle$. Both amplitude and 
phase noise have been included with $\Gamma_{\rm AD}=\Gamma_{\rm PD}=\Gamma$. Note the logarithmic scales for $\Gamma/\Delta$. We use the same parameters as in Fig.~\ref{fig3}. 
Panel (a): Fidelity $\overline{F_{|0\rangle}}$ vs $\Gamma/\Delta$, 
 averaged over $500$ measurement trajectories, for several values of $\Delta\delta t$. 
 The threshold, $\Gamma_c/\Delta\simeq 0.1$, is marked by an arrow. 
Panel (b): Same as panel (a) but for $\Delta\delta t=0.03$ and several values of $J/\Delta$.
Panel (c): Averaged fidelity (red curve) and variance (black bars) vs $\Gamma/\Delta$ for $\Delta\delta t=0.01$. The variance has been computed from 500 trajectories. 
}
\label{fig5}
\end{figure*}
\begin{figure}
\includegraphics[scale=0.5]{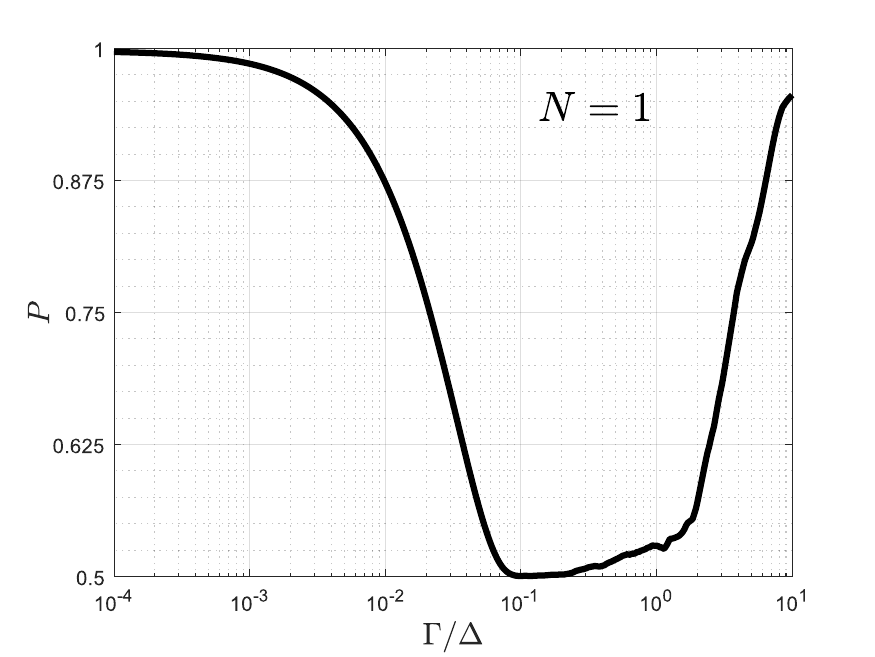}
\caption{Purity $P$ in Eq.~\eqref{purity} vs $\Gamma /\Delta$ for 
$N=1$, including amplitude and phase damping with $\Gamma_{\rm AD}=\Gamma_{\rm PD}=\Gamma$.
Note the logarithmic scales for $\Gamma/\Delta$.  
We use the same parameters as well as initial and target states as in Fig.~\ref{fig4}(a), 
with $\Delta \delta t=0.03$ and $\Lambda/\Delta=0.98$. 
Averages are over $500$ measurement trajectories.}
\label{fig6}
\end{figure}
In order to quantify the closeness of the time-evolving system state $\rho(t)$ to the (pure) target state $|\Psi_f\rangle$, a convenient measure is given by the standard fidelity \cite{Nielsen2000},
\begin{equation}\label{fidelity}
    F_{|\Psi_f\rangle}(t) = \langle \Psi_f| \rho(t) |\Psi_f\rangle.
\end{equation}
This real number is bounded by $0\le F(t) \le 1$, with $F=1$ only if $\rho(t)=|\Psi_f\rangle\langle \Psi_f|$ is fully converged.
When averaged over many measurement trajectories, we denote the averaged fidelity by $\overline{F_{|\Psi_f\rangle}(t)}$.  
Figure~\ref{fig4}(a) shows the time-dependent fidelities $F_{|0\rangle}(t)$ and $F_{|1\rangle}(t)$ 
for an individual measurement trajectory with $|\Psi(0)\rangle=|+\rangle$ and $|\Psi_f\rangle=|0\rangle$, where the active steering protocol is subject to very weak amplitude and phase damping,
$\Gamma_{\rm AD}=\Gamma_{\rm PD}=\Gamma$. 
We observe that apart from occasional and inevitable quantum jump events, the state is reliably steered towards the desired target state $|0\rangle$.  If a quantum jump (with $\xi=1$) occurs, it comes with a large change in the fidelity. However, the protocol ``repairs'' this glitch, resulting from an unlikely and unfavorable measurement outcome, quickly by itself again.  Our active steering protocol can thus be viewed as an implementation of  autonomous state stabilization.
In general, our simulation results show that the number of time steps needed for approaching the final state is $\sim (J\delta t)^{-1}$.

For the same $N=1$ protocol as just discussed,
 we next turn to the asymptotic long-time value of the averaged fidelity, 
$\overline{F_{|0\rangle}}$, as a function of the error rate $\Gamma$.  
We again consider the case where both amplitude and phase noise are present. 
In general, $\overline{F_{|0\rangle}}$ depends on $\delta t$  
for otherwise fixed parameters.  (Using $\Delta$ as reference energy, the dependence on $\delta t$ is encoded by the dimensionless parameter $\Delta \delta t$.) 
However, we observe from Fig.~\ref{fig5}(a) that
results for different $\delta t$ (approximately) fall onto a single scaling curve when plotting the fidelity as a function of $\Gamma/\Delta$.  The scaling curve is characterized by a sharp threshold error rate $\Gamma_c$ where a qualitative change happens.   For the present parameters, we find $\Gamma_c/\Delta \simeq 0.1$. In order to investigate how this critical value changes when changing physical model parameters for a fixed value of $\Delta\delta t$, Fig.~\ref{fig5}(b) shows how the curves $\overline{F}$ vs $\Gamma/\Delta$ change when $J/\Delta$ is varied. 
Remarkably,  both the threshold value $\Gamma_c$ and the scaling property
are basically unaffected by changing $J$ as long as we are in the regime $J\agt \Lambda$, where steering couplings
are significant and the protocol remains efficient.  However, the scaling feature of the fidelity now breaks down for strong damping, $\Gamma>\Gamma_c$, where the precise value of $J/\Delta$ matters.

For $\Gamma\ll \Gamma_c$, we conclude that amplitude and phase noise do not harm the active steering 
protocol. One can still achieve high fidelities, and it is possible to approach and/or stabilize the desired target state with high accuracy. However, with increasing error rate, upon approaching $\Gamma=\Gamma_c$, a qualitative and sudden change in the average fidelity occurs.  Apart from the fidelity, this threshold also shows up in other diagnostic measures as detailed below.  We note that Fig.~\ref{fig5}(c) shows the variance of the fidelity (represented by error bars). 
For $\Gamma\ll \Gamma_c$, the variance is small since basically only
the random measurement outcomes cause fidelity variations between different measurement trajectories.  As $\Gamma$ grows, however, also the variance increases since stochastic bath forces now become significant and cause a larger spread.

Next,  in Fig.~\ref{fig6}, we show the purity $P$ of the late-time averaged system state,
\begin{equation}\label{purity}
P = {\rm Tr}\left( \overline {\rho ( t \to \infty)}^2\right),
\end{equation}
as function of $\Gamma /\Delta$ for $N=1$. We use
the same parameters and the same initial and target states as before. 
We observe that the fidelity threshold at $\Gamma=\Gamma_c$ in Fig.~\ref{fig5} coincides with a \emph{purity gap closing} 
\cite{Bardyn2018}. For $\Gamma= \Gamma_c$, the asymptotic averaged state is given by the
infinite-temperature state, $\overline{\rho}=\frac12\sigma^0$, with minimal purity $P=1/2$. 
Purity gap closing transitions for mixed density matrices describing ensemble-averaged states have been described within a general theoretical framework in Ref.~\cite{Bardyn2018}.

On the other hand, one reaches the target state with high purity for $\Gamma\ll \Gamma_c$. 
In this case, errors remain correctable in our protocol. For $\Gamma> \Gamma_c$, however, the dynamics
is dominated by dissipation channels associated with $\Gamma$ in Eq.~\eqref{SME}. We observe that
one approaches a pure state again, see also Fig.~\ref{fig3}(b), where the purity $P$ 
increases towards its maximum value $P=1$ as $\Gamma$ increases.
For large $\Gamma$, this pure state is linked to the dark state of the dissipative Lindbladian 
instead of the target state $|\Psi_f\rangle$ which for high amplitude damping and dephasing is given by the ground state $|1\rangle$ of $H_s$.
Remarkably, our numerical simulations show that for $\Gamma\approx \Gamma_c$, the competition between the unitary time evolution due to steering operators and the measurement- and noise-induced dissipative processes results in an 
infinite-temperature state, where control over the system state is basically lost and the purity gap closes. 
We note that the error threshold has been discussed above for the ensemble averaged system, using both the fidelity and the purity as diagnostic measures.  
For individual measurement trajectories, signatures of the error threshold can also be observed but these are then subject to random fluctuations, see Fig.~\ref{fig5}(c).

In order to rationalize the existence of a threshold transition when varying the error rate $\Gamma$, 
let us consider the measurement-averaged state 
change $\langle d\rho \rangle_{\rm ms}$ in one time step, starting from the actual 
state $\rho=\rho(t)$.   
Neglecting $E_A\ll J$ in $H_s$ and writing the steering operator as $H_{s,\alpha}=\pm J\sigma^\alpha$, Eq.~\eqref{SME} yields 
\begin{eqnarray}\nonumber
\langle d\rho\rangle_{\rm ms} &=& \mp i J\delta t\, [\sigma^\alpha,\rho] + \delta t \Gamma_{\rm ms} \,
{\cal D}[\sigma^s] \rho \\ & +&  \delta t \Gamma_{\rm AD}\, {\cal D}(\sigma^-)\rho + \frac12\delta t \Gamma_{\rm PD}\,
{\cal D}[\sigma^z] \rho . \label{avlind}
\end{eqnarray}
Here three different energy scales appear,  namely $J$, $\Gamma$ (assuming $\Gamma_{\rm AD}=\Gamma_{\rm PD})$, and $\Gamma_{\rm ms}=\Lambda^2 \delta t$. The latter scale is the measurement-induced decay rate.
For $J\ll \Gamma_{\rm ms}$, a balance between measurement-induced and noise-induced terms is expected for $\Gamma_c \approx \Gamma_{\rm ms}$.  
However, this balance does not result in a purity gap closing transition. In fact, in this limit, our numerical simulations do not find a sharp threshold transition as
described above. For the case $J\gg \Gamma_{\rm ms}$ considered above, on the other hand, the existence of the error threshold is robust under changes of $J$ and $\Lambda$.  In this regime, the steering dynamics is clearly important, 
and one has to consider the competition between the unitary term and the damping terms (due to $\Gamma_{\rm ms}$ and $\Gamma$) as well as the active decision making.

From Eq.~\eqref{avlind}, assuming $O_s=\sigma^z$, we can analytically compute the anticipated average change $\langle (dO_s)_\alpha\rangle_{\rm ms}={\rm Tr} (\langle d\rho\rangle_{\rm ms}\, O_s)$ from Eqs.~\eqref{dos} and \eqref{avlind}. Writing $\rho(t)=\frac12(\sigma^0+{\bf r}\cdot \boldsymbol{\sigma})$ in terms of a Bloch vector ${\bf r}={\bf r}(t)=(r_x,r_y,r_z)^T$, we find 
\begin{eqnarray}
\langle (dO_s)_\alpha\rangle_{\rm ms} &=& \pm 2 J\delta t \sum_{j} \varepsilon_{z\alpha j} r_j - \delta t\, \Gamma_{\rm AD} (1+r_z)\\ \nonumber 
&-& \delta t \Gamma_{\rm ms} \Biggl[ 2\left(1-\frac{\cos^2(\phi_0/2)}{1-{\cal T}\sin^2(\phi_0/2)}\right) r_z \\
&+& \frac{\sqrt{1-{\cal T}}\, \sin\phi_0}{1-{\cal T}\sin^2 (\phi_0/2)} r_y \Biggr ]\nonumber
\end{eqnarray}
with the antisymmetric $\varepsilon$-tensor.
We note that the phase noise rate $\Gamma_{\rm PD}$ does not appear here when assuming the target state $|0\rangle$.  
In any case, only the first term depends on the steering operator choice, so the active decision making will always choose either $\alpha=x$ or $\alpha=y$, depending on whether $|r_x|>|r_y|$ or vice versa, with the sign 
determined by the sign of $r_j$ with $j\ne (\alpha,z)$.  For $r_x=r_y=0$, the decision making is frustrated: any steering operator leads to the same choice.

Depending on the measurement outcome $\xi\in \{0,1\}$, the SME \eqref{SME} becomes an iteration equation for the Bloch vector, ${\bf r}_\xi(t+\delta t)-{\bf r}(t)=d{\bf r}_\xi$.
Including the effects of $E_A$ in the SME, for the steering operator $\pm J\sigma^\alpha$, we then find (we use the summation convention and $j\in \{1,2,3\}=\{x,y,z\}$)
\begin{eqnarray}\label{SME3}
  d r_{j,\xi}   &=& {\rm Tr}(\sigma^j d\rho_\xi) =  
  2 \delta t\, r_k [E_A \varepsilon_{jzk} \pm J \varepsilon_{j\alpha k}] \\ \nonumber
  &+& \frac12 \xi \, r_k {\rm Tr}[\sigma^j\sigma^s\sigma^k\sigma^s-\sigma^j\sigma^k] \\ \nonumber
  &+& \delta t \,\Gamma_{\rm AD} {\rm Tr}\left ( \sigma^j [\sigma^-\rho\sigma^+ - \frac12 \{\sigma^+\sigma^-,\rho\}] \right) \\
  \nonumber
  &+& \delta t \, \Gamma_{\rm PD} r_j ( \delta_{j,z}- 1 ) . 
\end{eqnarray}

For simplicity, let us now study the special
parameter choice $\phi_0=\pi$, ${\cal T}<1$, and $\Gamma_{\rm AD}=0$,  where
 the above equations yield
\begin{equation}
\langle (dO_s)_\alpha\rangle_{\rm ms} = 2\delta t \, {\bf r}\cdot \left(\begin{array}{c} \mp J\delta_{\alpha,y} \\
\pm J\delta_{\alpha,x} \\ -\Gamma_{\rm ms} \end{array} \right),   
\end{equation}
and from Eq.~\eqref{SME3}
\begin{eqnarray}\nonumber
d{\bf r}_\xi &=& 2\delta t  \left( \Delta\sqrt{1-{\cal T}}\, \hat e_z\times {\bf r} \pm J \hat e_\alpha \times {\bf r} +
\frac{\Gamma_{\rm PD}}{2} [ r_z \hat e_z - {\bf r}] \right) \\
&-& 2\xi({\bf r}-r_y \hat e_y), \label{drxi}
\end{eqnarray}
 where $\hat e_\alpha$ is a unit vector in direction $\alpha$.
We recall that here the effects of $E_A$, see Eq.~\eqref{ABS}, have been taken into account. 
We also note that a numerical implementation of the dynamics in this representation is  unstable because one has to ensure $|{\bf r}(t)|\le 1$ at all times. For this reason, we have adopted the SSE discussed in Sec.~\ref{sec2c} for the above
numerical analysis. If we ignore quantum jumps (i.e., we put $\xi=0$) and assume ${\bf r}=(r_x<0,0,r_z>0)^T$ for simplicity, we see that $H_{s,\alpha}=-J\sigma^y$ will be chosen by the active decision making.  Equation \eqref{drxi} then implies
\begin{eqnarray}\nonumber
    d{\bf r}_{\xi=0} &=& -2\delta t  \Bigl( \Delta\sqrt{1-{\cal T}} r_y+J r_z+ \frac{1}{2}\Gamma_{\rm PD}r_x\\ &-&\Delta\sqrt{1-{\cal T}} r_x+\frac{1}{2}\Gamma_{\rm PD} r_y - J r_x \Bigr).
\end{eqnarray}
This expression predicts an approximate stationary point at $r_x=-r_z$ for $\Gamma_{\rm PD}\simeq 2J$. The $y$-component
will then drive this state toward ${\bf r}=0$, i.e., toward the infinite temperature state characterizing the threshold value of $\Gamma_{\rm PD}$.  
However, the above argument ignores quantum jumps, which can modify the actual value of the threshold rate.  While the numerically observed critical value $\Gamma_c$ is found to be roughly one order of magnitude smaller than $J$, these considerations are consistent with the existence of a steering-induced threshold transition. 

 \begin{figure*}
\includegraphics[scale=0.5]{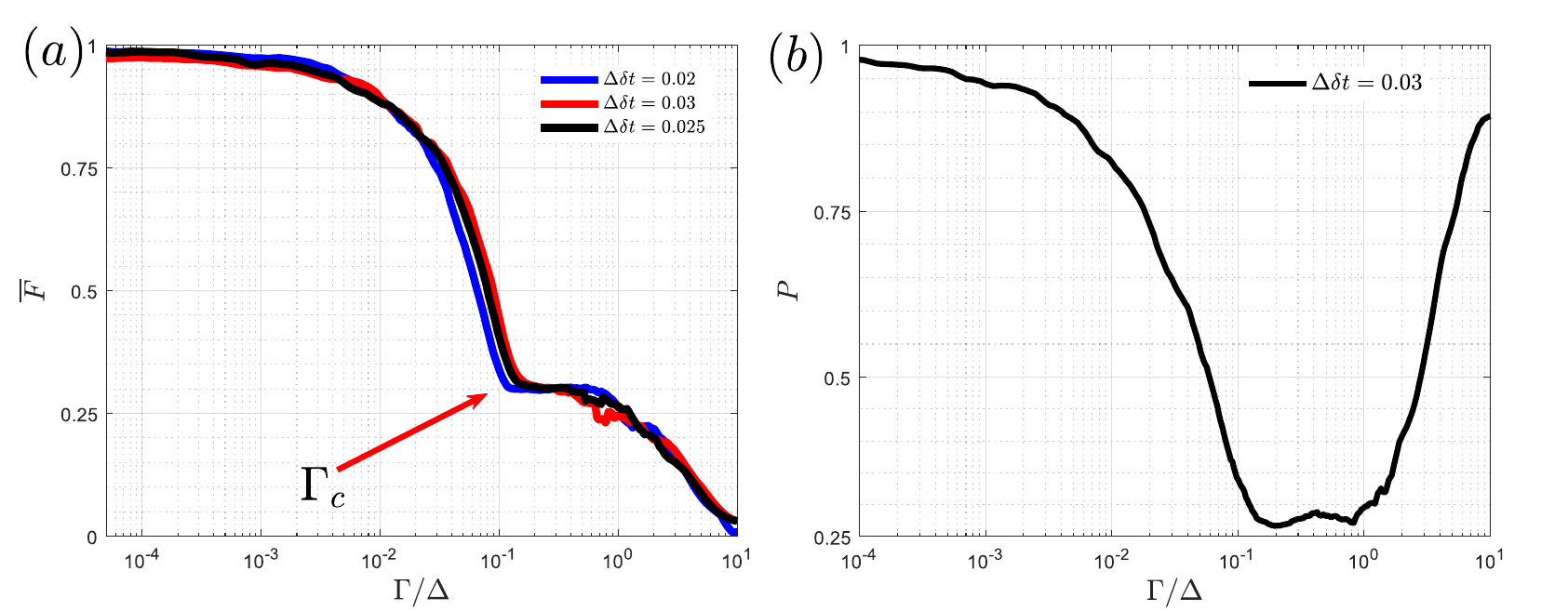}
\caption{Active steering results for $N=2$, with the same parameters as in Fig.~\ref{fig3} for 
 both system-plus-detector subsystems but using $\Lambda/\Delta =0.49$ and different 
  $\Delta \delta t$. We include both amplitude and phase noise, $\Gamma_{\rm AD}=\Gamma_{\rm PD}=\Gamma$. 
 Note the logarithmic scales for $\Gamma/\Delta$. Results are averaged over 500 measurement trajectories.
Panel (a): Averaged late-time fidelity $\overline{F_{|\Phi_{0,+}\rangle}}$ vs $\Gamma/\Delta$ for 
 $|\Psi(0)\rangle=|++\rangle$ and the Bell target state  $|\Psi_f\rangle=|\Phi_{0,+}\rangle$.  
 The threshold value $\Gamma_c$ is marked by an arrow.
 Panel (b):  Purity $P$ vs $\Gamma/\Delta$ for $\Delta \delta t=0.03$.    } 
\label{fig7}
\end{figure*}

\begin{figure}
\includegraphics[scale=0.55]{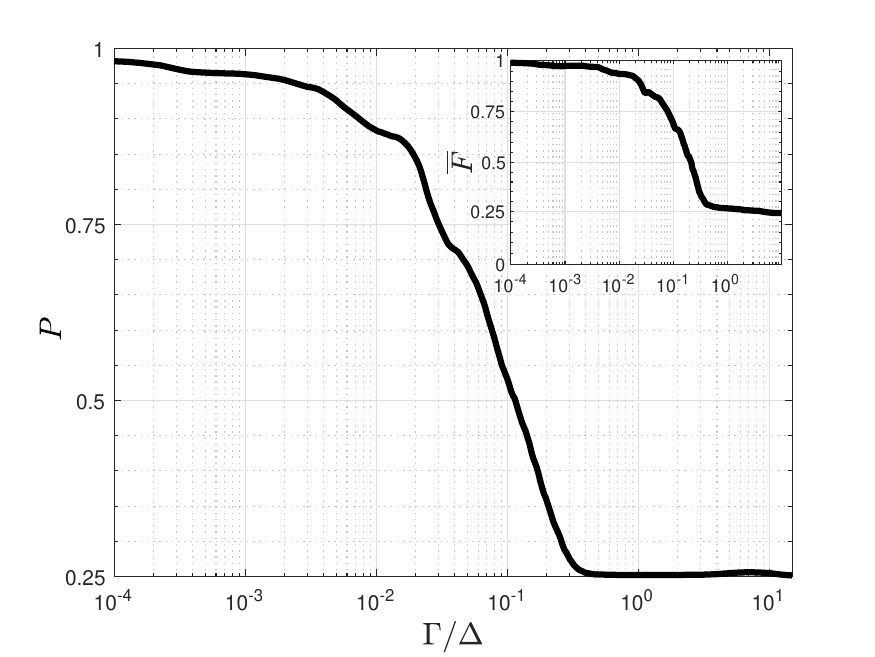}
\caption{Simulation results for the late-time state reached by the $N=2$ active steering protocol in Fig.~\ref{fig7}
if only phase noise is present, $\Gamma_{\rm AD}=0$ and $\Gamma_{\rm PD}=\Gamma$. Note the logarithmic scales for $\Gamma/\Delta$. Averages were taken over $1000$ measurement trajectories, 
using $\Delta \delta t=0.03$ and $\Lambda/\Delta =0.49$.
Main panel: Purity $P$ vs $\Gamma/\Delta$, see Eq.~\eqref{purity}.
Inset: Averaged fidelity $\overline{F_{|\Phi_{0,+}\rangle}}$ vs $\Gamma/\Delta$.  
}
\label{fig8}
\end{figure}

\subsection{Two system qubits}\label{sec3b}

We next analyze the case of $N=2$ system qubits using the protocol described in Sec.~\ref{sec2b}, 
where we assume identical parameters for both system-plus-detector subsystems in Fig.~\ref{fig1}, with $J_1=J_2=J.$ 
In Sec.~\ref{sec4}, we briefly comment on what happens for asymmetric parameter choices.  
In principle, we now have $7^2=49$ steering operators at hand, see Sec.~\ref{sec2b}. For specific tasks, it may be possible to reduce the set of steering operators to a smaller subset that 
still allows for reaching the desired target state.  While we have not explored this option systematically,
for setups with $N>2$, it will be important to determine the minimal set of steering  operators
in order to obtain efficient protocols.  

Again assuming that both amplitude and phase noise are present, $\Gamma_{\rm AD}=\Gamma_{\rm PD}=\Gamma$, 
Fig.~\ref{fig4}(b) shows that for weak error rate $\Gamma$, the $N=2$ protocol is able to converge the time-evolving state $\rho(t)$ with high fidelity towards a predesignated target state $|\Psi_f\rangle$.  In this specific example, we start from the initial state $|\Psi(0)\rangle=|++\rangle$ and steer towards the symmetric Bell target state, $|\Psi_f\rangle=|\Phi_{0,+}\rangle$, see Eq.~\eqref{bell2}.  
We observe that the protocol is able to retain high fidelity for arbitrarily long times, such that
the state is autonomously stabilized. For an individual measurement trajectory, see Fig.~\ref{fig4}(b), small-amplitude Rabi oscillations can be resolved in the time-dependent fidelity.  
The frequency of these oscillations corresponds to the energy difference $2E_A$
between the system qubit states, see Eq.~\eqref{ABS}.  As for the $N=1$ case in Fig.~\ref{fig4}(a),
quantum jumps occasionally occur and reduce the fidelity for a very short time span.  
However, the fidelity recovers quickly to a high value again, demonstrating the robustness of the autonomous state stabilization protocol. In fact, for weak damping, we find that the $N=2$ protocol is more efficient than its $N=1$ variant because of the state-purifying capabilities of the non-unitary jump operators for $N=2$.

For the same $N=2$ protocol, we show the averaged fidelity reached at late times as function of
$\Gamma /\Delta$ in Fig.~\ref{fig7}(a).  Similar to the $N=1$ case in Fig.~\ref{fig5}, we again
find a sharply defined error threshold, $\Gamma=\Gamma_c$, separating a weak-damping regime ($\Gamma<\Gamma_c$) 
and a strong-damping regime where steering fails ($\Gamma>\Gamma_c$).  
The threshold is again accompanied by a closure of the purity gap, as shown in Fig.~\ref{fig7}(b). 
For the considered parameters, the threshold error rate is $\Gamma_c/\Delta \simeq 0.15$.  
For $\Gamma=\Gamma_c$, we have an infinite-temperature state, 
$\overline{\rho}=\frac{1}{4}\sigma_1^0\sigma_2^0$, with minimal purity $P=1/4$.  
For $\Gamma\ll \Gamma_c$, the averaged state approaches
the pure target state  $|\Psi_f\rangle$ with high fidelity.  
By analogy to the QEC case, one may expect that in the limit of large $N$, the fidelity and the purity approach
their ideal values for all $\Gamma<\Gamma_c$.  However, for $N=2$, the fidelity and the purity are gradually degraded
upon increasing $\Gamma$ towards $\Gamma_c$.
For strong damping $\Gamma\gg\Gamma_c$, on the other hand, one approaches a pure dark state associated with the dissipator $\propto \Gamma$ in Eq.~\eqref{SME2}.  
As seen in Fig.~\ref{fig7}(a), the scaling property (i.e., the averaged fidelity depends on $\Gamma/\Delta$ without explicit dependence on $\delta t$) holds again.

We next consider the same $N=2$ protocol but in the presence of only phase noise, 
$\Gamma_{\rm AD}=0$ and $\Gamma_{\rm PD}=\Gamma$.  The main difference to the previously studied case is that the 
late-time state reached at large $\Gamma$ is now the infinite-temperature state.  One therefore
expects that for $\Gamma>\Gamma_c$, this minimal-purity state persists for all $\Gamma$.
As shown in Fig.~\ref{fig8}, this is  confirmed by our numerical simulation results.  For the parameters in Fig.~\ref{fig8}, we find $\Gamma_c/\Delta\simeq 0.3$.
For $\Gamma>\Gamma_c$,  the state is unsteerable and characterized by a very low and approximately constant fidelity (see inset) as well as minimal purity (main panel).  

To summarize, the sharp threshold separating a weak-damping regime ($\Gamma<\Gamma_c$, where errors remain correctable) 
from a strong-damping regime ($\Gamma>\Gamma_c$, where one cannot steer to the desired target
state anymore) is also observed in our $N=2$ simulations. 
Right at the threshold, the purity gap closes and one has an infinite-temperature state.  
For strong damping rates, one in general approaches a pure dark state dominated by the error channels.
If only phase noise is present, one instead finds an infinite-temperature state for all $\Gamma>\Gamma_c$.

\section{Conclusions and Outlook}\label{sec4}

In this work, we have formulated active steering protocols in the presence of error channels for few-qubit systems.
Our numerical simulations for $N=1$ and $N=2$ system qubits reveal a purity gap closing threshold 
at a specific error rate $\Gamma_c$ in such an autonomous state stabilization protocol.  
For $\Gamma\ll \Gamma_c$, the protocol steers the system to the predesignated target state with high fidelity and 
high purity.  For $\Gamma>\Gamma_c$, it is not possible to reach the target state anymore, and for $\Gamma \gg \Gamma_c$, the (undesired) stationary state of the pure error dynamics is instead attained.  The competition between the two dynamical resources -- active steering and errors -- each separately stabilizing a different stationary state,  leads to a  transition which is found to be sharp, 
and accompanied by a closure of the purity gap.  We expect that such thresholds are likely to appear also in larger-$N$ variants of our protocol. We have presented results for amplitude and phase noise with equal rates, $\Gamma_{\rm AD}=\Gamma_{\rm PD}$, 
and for phase noise only, $\Gamma_{\rm AD}=0$.  We note in passing that for $\Gamma_{\rm PD}=0$,  i.e., if only amplitude noise is present, the numerically observed threshold behavior (not shown here) is observed to be qualitatively similar as for $\Gamma_{\rm AD}=\Gamma_{\rm PD}$. 

It is worth noting that feedback control of a single qubit based on continuous measurements has a long history in quantum optics and quantum information science, see, e.g., Refs.
\cite{Audretsch2002,Wiseman2010,Uys2018}.  Similarly, purification protocols for single qubit states are specific examples for such protocols for this type of feedback control, where the measurement
basis changes through the course of the time evolution in order to maximize the state purity.  Beyond the single-qubit case, certain multi-qubit entangled states can also be stabilized via feedback based
on the measurement of collective system observables, see, e.g., Ref.~\cite{Stockton2004}.  In the present case, a weak measurement variant of entanglement swapping has been employed for two system qubits to achieve this goal.

The approach presented here is based on an active steering protocol where error channels are accounted for within
the SME. A cost function maximizing the expectation value of a physical observable 
has been used to guide the measurement trajectory. Since the present formulation requires that one keeps track 
of the system state, it is not scalable and restricted to relatively small values of $N$, see also Ref.~\cite{Morales2024}.  
The state tracking requirement is a well-known and general obstacle for many state-based feedback schemes. However, for small and very well characterized systems,
it may be possible to formulate an active steering protocol only in terms of  measurable quantities. In such cases, feedback protocols could avoid state tracking.  A detailed study of this interesting question is, however, beyond the scope of this paper. 

We note that for the $N=2$ case in Sec.~\ref{sec3b}, we have assumed identical parameters for both system-plus-detector subsystems. To ensure that our approach is robust when relaxing this assumption, we have performed additional simulations with parameter asymmetries of order $\sim 3\%$.  The corresponding results do not show 
qualitative differences to the results reported in Sec.~\ref{sec3b}, indicating that the 
purity gap closing threshold transition is robust. 

For future work, it would be interesting to study circuits with $N>2$ qubits, see also Ref.~\cite{Morales2024}.  
However, the computational cost of the present scheme grows exponentially with $N$, both due to the increase in Hilbert space size and in the number of steering operators.  On the other hand, the fact that for weak damping, the $N=2$ scheme is superior to the $N=1$ approach, together with the option of using a smaller set of steering operators,  suggests that one may be able to study systems with, say, $N=3$ or $N=4$, with only minor modifications.   
An important question for future research is to develop scalable variants of our protocol.  Once available, one could study
if there is a true phase transition associated with the threshold rate $\Gamma_c$ in the limit $N\to \infty$.

In this work, we have not studied the effects of errors that can induce leakage out of the computational qubit space, e.g., excitations to higher energy levels or quasiparticle poisoning effects.  It will be interesting to clarify in future research whether such errors are also associated with a threshold. In Refs.~\cite{Zazunov2014,Ackermann2023}, the Andreev state dynamics has been studied allowing for parity transitions. Incorporating such transitions into our protocol as an additional error channel could be the object of future work.
A further intriguing direction aims to generalize the target state from a single state (i.e. a single point on the Bloch sphere) to a two-dimensional state manifold defining a general qubit in order to exhibit control over the whole qubit space. This will provide an active steering variant of quantum error correction protocols \cite{Lieu2020,Shtanko2023,Kristensen2023}. Such protocols are not restricted to stabilizer codes.

All data underlying the figures presented in this work can be retrieved at zenodo:
\textcolor{blue}{https://zenodo.org/records/11581356}.\\

\begin{acknowledgments} 
We thank M. Devoret, Y. Gefen, I. Gornyi, C. Sch\"onenberger, A. Zazunov, and especially M. M\"uller for discussions and
comments on the manuscript.
We acknowledge funding by the Deutsche Forschungsgemeinschaft (DFG, German Research Foundation) under Projektnummer 277101999 - TRR 183 (projects B02 and C01), under Project No.~EG 96/13-1, and under Germany's Excellence Strategy - Cluster of Excellence Matter and Light for Quantum Computing (ML4Q) EXC 2004/1 - 390534769, and from the Spanish Ministerio de Ciencia e Innovaci\'on via grants TED2021-130292B-C41, and through the ``Mar\'{i}a de Maeztu'' Programme for Units of Excellence in R{\&}D CEX2023-001316-M. 
\end{acknowledgments}

\bibliography{biblio}

\end{document}

%% file: f1.tex
	 \thispagestyle{empty}
	\begin{circuitikz}[scale=0.85]
		\ctikzset{
			capacitors/scale=0.6,
			inductors/coils=5
		}
		
		\begin{scope}[yshift=0cm, xshift=2.5cm]
			
			\tikzset{meter/.append style={draw, inner sep=2.5, rectangle, font=\vphantom{A}, minimum width=15, minimum height=10, line width=.8, fill=green!20,
				path picture={
					\draw[black] ([shift={(2pt,4pt)}]path picture bounding box.south west) to[bend left=50] ([shift={(-2pt,4pt)}]path picture bounding box.south east);
					\draw[black,-latex] ([shift={(0,2pt)}]path picture bounding box.south) -- ([shift={(5pt,-1pt)}]path picture bounding box.north);
		}}}
			
			\draw [line width=1pt, fill=orange!60] (1.4,-3.2)
			to (0.7,-3.2)
			to(0.7,-2.5)
			to (1.4,-2.5) -- (1.4,-3.2);
			\node at (1.05,-2.85) {$s_1$};
			
			\draw [line width=1pt, fill=cyan!30] (1.05,-1.5) circle (0.42cm);
			\node at (1.05,-1.5) {$d_1$};
			
			\draw [line width=1pt][dashed] (1.05,-1.95) -- (1.05,-2.45);

			\draw [line width=1pt, fill=orange!60] (2.4,-3.2)
			to (1.7,-3.2)
			to(1.7,-2.5)
			to (2.4,-2.5) -- (2.4,-3.2);
			\node at (2.05,-2.85) {$s_2$};
			
			\draw [line width=1pt, fill=cyan!30] (2.05,-1.5) circle (0.42cm);
			\node at (2.05,-1.5) {$d_2$};
			
			\draw [line width=1pt][dashed] (2.05,-1.95) -- (2.05,-2.45);

			\draw [line width=1pt] (0.65,-1.12) .. controls (1.05,-0.88) and (2.05,-0.88) .. (2.45,-1.12);
			
			\node[meter] at (1.55,-0.65) {};
			
			\draw[->, line width=1pt] (2.8,-2.2) -- (3.3,-2.2);
			
			\draw [line width=1pt, fill=black!10] (3.35,-3.) -- (6.1,-3.) -- (6.1,-1.7) -- (3.35,-1.7) -- cycle;
			
		\end{scope}
		
		\begin{scope}[yshift=-3cm, xshift=2.5cm]
			
			\tikzset{meter/.append style={draw, inner sep=2.5, rectangle, font=\vphantom{A}, minimum width=15, minimum height=10, line width=.8, fill=green!20,
				path picture={
					\draw[black] ([shift={(2pt,4pt)}]path picture bounding box.south west) to[bend left=50] ([shift={(-2pt,4pt)}]path picture bounding box.south east);
					\draw[black,-latex] ([shift={(0,2pt)}]path picture bounding box.south) -- ([shift={(5pt,-1pt)}]path picture bounding box.north);
		}}}
			
			\draw [line width=1pt, fill=orange!60] (1.4,-3.2)
			to (0.7,-3.2)
			to(0.7,-2.5)
			to (1.4,-2.5) -- (1.4,-3.2);
			\node at (1.05,-2.85) {$s_1$};
			
			\draw [line width=1pt, fill=cyan!30] (1.05,-1.5) circle (0.42cm);
			\node at (1.05,-1.5) {$d_1$};
			
			\draw [line width=1pt][dashed] (1.05,-1.95) -- (1.05,-2.45);

			\draw [line width=1pt, fill=orange!60] (2.4,-3.2)
			to (1.7,-3.2)
			to(1.7,-2.5)
			to (2.4,-2.5) -- (2.4,-3.2);
			\node at (2.05,-2.85) {$s_2$};
			
			\draw [line width=1pt, fill=cyan!30] (2.05,-1.5) circle (0.42cm);
			\node at (2.05,-1.5) {$d_2$};
			
			\draw [line width=1pt][dashed] (2.05,-1.95) -- (2.05,-2.45);

			\draw [line width=1pt] (0.65,-1.12) .. controls (1.05,-0.88) and (2.05,-0.88) .. (2.45,-1.12);
			
			\node[meter] at (1.55,-0.65) {};
			
			\draw[->, line width=1pt] (2.8,-2.2) -- (3.3,-2.2);
			
			\draw [line width=1pt, fill=black!10] (3.35,-3.) -- (6.1,-3.) -- (6.1,-1.7) -- (3.35,-1.7) -- cycle;

		\end{scope}
	
	\begin{scope}[yshift=0, xshift=0.4cm]

		\tikzset{meter/.append style={draw, inner sep=2.5, rectangle, font=\vphantom{A}, minimum width=15, minimum height=10, line width=.8, fill=green!20,
				path picture={
					\draw[black] ([shift={(2pt,4pt)}]path picture bounding box.south west) to[bend left=50] ([shift={(-2pt,4pt)}]path picture bounding box.south east);
					\draw[black,-latex] ([shift={(0,2pt)}]path picture bounding box.south) -- ([shift={(5pt,-1pt)}]path picture bounding box.north);
		}}}
		\draw [line width=1pt, fill=orange!60] (1.4,-3.2)
		to (0.7,-3.2)
		to(0.7,-2.5)
		to (1.4,-2.5) -- (1.4,-3.2);
		\node at (1.05,-2.85) {s};
		
		\draw [line width=1pt, fill=cyan!30] (1.05,-1.5) circle (0.42cm);
		\node at (1.05,-1.5) {d};
		
		\draw [line width=1pt][dashed] (1.05,-1.95) -- (1.05,-2.45);
		
		\draw [line width=1pt] (0.6,-1.1) .. controls (0.9,-0.88) and (1.15,-0.88) .. (1.45,-1.1);
		
		\node[meter] at (1.05,-0.65) {};
	\end{scope}

	\begin{scope}[yshift=-3cm, xshift=0.4cm]

	\tikzset{meter/.append style={draw, inner sep=2.5, rectangle, font=\vphantom{A}, minimum width=15, minimum height=10, line width=.8, fill=green!20,
				path picture={
					\draw[black] ([shift={(2pt,4pt)}]path picture bounding box.south west) to[bend left=50] ([shift={(-2pt,4pt)}]path picture bounding box.south east);
					\draw[black,-latex] ([shift={(0,2pt)}]path picture bounding box.south) -- ([shift={(5pt,-1pt)}]path picture bounding box.north);
		}}}
	\draw [line width=1pt, fill=orange!60] (1.4,-3.2)
	to (0.7,-3.2)
	to(0.7,-2.5)
	to (1.4,-2.5) -- (1.4,-3.2);
	\node at (1.05,-2.85) {s};
	
	\draw [line width=1pt, fill=cyan!30] (1.05,-1.5) circle (0.42cm);
	\node at (1.05,-1.5) {d};
	
	\draw [line width=1pt][dashed] (1.05,-1.95) -- (1.05,-2.45);
	
	\draw [line width=1pt] (0.6,-1.1) .. controls (0.9,-0.88) and (1.15,-0.88) .. (1.45,-1.1);
	
	\node[meter] at (1.05,-0.65) {};
\end{scope}
		
		\draw[->, line width=1pt] (0.2,-6.7) -- (0.2,0.3);
		\draw[->, line width=1pt] (0.2,-6.7) -- (5.2,-6.7);
		\node at (-0.7,-5.2) {\large \(t=0\)};
		\node at (-0.7,-2.2) {\large \(t=\delta t\)};
		\node at (0.65,-4.5) {\normalsize \(|0\rangle_d\)};
		\node at (0.65,-1.5) {\normalsize \(|0\rangle_d\)};
		\node at (2.65,-4.5) {\normalsize \(|00\rangle_d\)};
		\node at (2.65,-1.5) {\normalsize \(|00\rangle_d\)};
		\node at (1.3,-7.2) {\large \(N=1\)};
		\node at (4.1,-7.2) {\large \(N=2\)};
		\node at (0.6,-0.2) {\large \((a)\)};
		\node at (3,-0.2) {\large \((b)\)};
		\node at (-0.2,-0.2) {\Huge \(\vdots\)};
		\node at (7.2,-6.4) {\large \( \text{Control Unit}\)};
		\node at (7.2,-5.05) {\footnotesize \( \text{calc.} \langle (dO_s)_\alpha\rangle_{\rm ms} \)};
		\node at (7.14,-5.4) {\footnotesize \( \rightarrow \text{choose } H_{s}\)};
		\node at (7.14,-5.75) {\footnotesize \( \text{for } t+\delta t\)};
		\node at (7.2,-3.4) {\large \( \text{Control Unit}\)};
		\node at (7.2,-2.05) {\footnotesize \( \text{calc.} \langle (dO_s)_\alpha\rangle_{\rm ms} \)};
		\node at (7.14,-2.4) {\footnotesize \( \rightarrow \text{choose } H_{s}\)};
		\node at (7.14,-2.75) {\footnotesize \( \text{for } t+2\delta t\)};
	\end{circuitikz}
	